# On the Use of Deep Learning in Software Defect Prediction

Görkem Giray[1]
gorkemgiray@gmail.com
0000-0002-7023-9469

Kwabena Ebo Bennin[2]
kwabena.bennin@wur.nl
0000-0001-9140-9271

Ömer Köksal[3]
koksal@aselsan.com.tr
0000-0003-1372-7033

Önder Babur[2,4]
onder.babur@wur.nl
0000-0002-1460-2825

Bedir Tekinerdogan[2]
bedir.tekinerdogan@wur.nl
0000-0002-8538-7261

[1] Independent Researcher, Izmir, Turkey
[2] Information Technology Group, Wageningen University & Research, Wageningen, The Netherlands
[3] ASELSAN, Ankara, Turkey
[4] Department of Mathematics and Computer Science, Eindhoven University of Technology, Eindhoven, The Netherlands

## Abstract

*Context:* Automated software defect prediction (SDP) methods are increasingly applied, often with the use of machine learning (ML) techniques. Yet, the existing ML-based approaches require manually extracted features, which are cumbersome, time consuming and hardly capture the semantic information reported in bug reporting tools. Deep learning (DL) techniques provide practitioners with the opportunities to automatically extract and learn from more complex and high-dimensional data.

*Objective:* The purpose of this study is to systematically identify, analyze, summarize, and synthesize the current state of the utilization of DL algorithms for SDP in the literature.

*Method:* We systematically selected a pool of 102 peer-reviewed studies and then conducted a quantitative and qualitative analysis using the data extracted from these studies.

*Results:* Main highlights include: (1) most studies applied supervised DL; (2) two third of the studies used metrics as an input to DL algorithms; (3) Convolutional Neural Network is the most frequently used DL algorithm.

*Conclusion:* Based on our findings, we propose to (1) develop more comprehensive DL approaches that automatically capture the needed features; (2) use diverse software artifacts other than source code; (3) adopt data augmentation techniques to tackle the class imbalance problem; (4) publish replication packages.

**Acronyms:**
SE: Software Engineering
ML: Machine Learning
DL: Deep Learning
SDP: Software Defect Prediction
WPDP: Within Project Defect Prediction
CPDP: Cross Project Defect Prediction
HDP: Heterogeneous Defect Prediction
SMS: Systematic Mapping Study
SLR: Systematic Literature Review
RQ: Research Question

## TABLE OF CONTENTS

## 1 INTRODUCTION

Software reliability and quality mainly depend on removing faults or defects in software. Although some defects might arise from causes unrelated to code (such as compilers or byte code representations), the main source of software faults is software code. The traditional way of finding software defects is by testing and conducting reviews. However, these activities may require extensive time and effort. On the other hand, automatic prediction of defective software modules at early stages may guide developers in improving code quality at a reduced cost compared to a fully manual approach (Wahono, 2015). To this end, software defect prediction (SDP) aims to promptly identify potential faults in the software and is a promising approach to improving software quality (Lessmann et al. 2008, Menzies et al., 2007). Therefore, SDP has become an important research topic in software engineering and testing in recent years.

Predicting defect-prone parts of software before discovering faults by performing substantial efforts is a challenging task. The main challenge of SDP is identifying the faulty parts of source code with better fault prediction performance. To this end, diverse methods and techniques have been proposed and reported in the literature for many years. Many researchers use learning-based algorithms to have better accuracies in SDP; on the other hand, some research has focused on the semantic representation of the source code. Researchers have been using machine learning (ML) and, more recently, deep learning (DL) algorithms to develop efficient SDP models. ML-based SDP techniques require manual extraction of features mainly based on software metrics. Although software metrics are effective indicators of defective portions of software (Rodríguez et al., 2012), manually extracted features are time-consuming to construct in the first place, and they hardly capture semantic information reported in bug reporting tools. On the other hand, DL-based techniques automatically extract higher-level features and learn from more complex and high-dimensional data (LeCun et al., 2015). Therefore, many researchers have recently focused on developing SDP models using DL-based techniques.

DL covers extensive state-of-the-art techniques and algorithms. Therefore, many studies applied these algorithms and techniques to the SDP domain comparing their findings with other studies. Hence, this paper systematically identifies, analyzes, summarizes, and synthesizes the current state of developing SDP models using DL algorithms and techniques. We conducted our survey using systematic literature review (SLR), a well-defined method introduced by Kitchenham and Charters (2007). Further, we indicate and investigate the recent research trends and point out future research directions on SDP. To this end, we selected a pool of 102 peer-reviewed studies and then conducted a quantitative and qualitative analysis using the data extracted from these studies. Researchers and practitioners may benefit from this survey to understand the state-of-the-art DL usage for SDP and shape their efforts to build more effective and efficient SDP models.

The rest of the paper is organized as follows: Section 2 provides a background and the related work. Section 3 explains the research methodology. Section 4 presents the results. Section 5 discusses our findings and reports the threats to validity. Section 6 concludes the paper.

## 2 BACKGROUND AND RELATED WORK

### 2.1 SOFTWARE DEFECT PREDICTION

SDP mainly involves prediction models that are built to predict faulty parts of software. Although diverse techniques and algorithms have been applied in order to have better performing (e.g., more accurate) SDP models, the main steps of SDP can be summarized as in Figure 1: (1) collect clean and defective code samples from software repositories, (2) extract features to create a dataset, (3) balance the dataset if it is imbalanced, (4) train a prediction model on the dataset, (5) predict the faulty parts for a dataset extracted from a new software (different version of trained dataset or new software project), and finally, (6) evaluate the performance of the SDP model. This process is iterative; Figure 1 ignores iteration steps for the sake of simplicity.

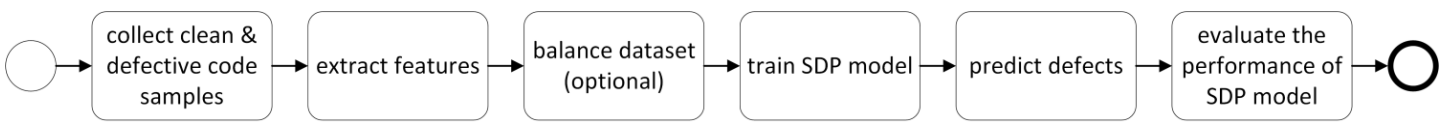

**Figure 1. Software defect prediction process**

Figure 1 shows that the process starts with collecting clean and defective code samples. Software data can be found in different formats comprising of source codes, commit messages, bug reports and other software artifacts. These data are usually extracted from archives and repositories.

The second step in SDP is feature extraction. During this phase, the software artifacts/source codes/commit logs and messages among others are converted to metrics which is used as input data for model training. The input data type, ranging from McCabe metrics (McCabe, 1976), CK metrics (Chidamber & Kemerer, 1994), change histories, assembly code, to source code, as well as the representation of the data are crucial in the feature extraction step. Besides metric-based data, nowadays, several DL techniques provide automatic extraction of features and learn from more complex and high dimensional data. In many studies in the literature, defect data from popular public defect repositories, such as the NASA (Shepperd et al., 2013) and PROMISE (Jureczko & Madeyski, 2010) datasets, have been utilized.

The next step is usually an optional step. This phase involves balancing the data since defect datasets typically include much fewer faulty parts than non-faulty. Unfortunately, most SDP techniques suffer from this class imbalance problem as several metrics for evaluating SDP performance result in misleading results due to the imbalanced structure of classes (Bennin et al., 2016). A variety of techniques, such as oversampling, can be used to tackle this issue and increase SDP performance.

The fourth step of SDP is determining the faulty parts of the software. The main concern in this step is the selection of DL algorithms and techniques, which can involve a wide range of architectures (e.g., Convolutional Neural Networks) and machine learning categories (e.g., supervised or not). In addition, the granularity of the faulty parts to be detected is an important issue at this step: these can be e.g., at module, file, class, function, or sentence level.

The next step is predicting the faulty parts of new (test) data using the trained model in the previous step. The prediction performed here provides the input for the last step of the SDP process.

The final step of the SDP process is evaluating the model developed. The SDP model can be evaluated utilizing various metrics such as F-measure or area under curve. One or more of such metrics are used to evaluate the prediction models and compare them with other related studies.

Orthogonal to the variety of choices in several steps of the process as outlined above, SDP studies can also be categorized with respect to their scenarios. Traditionally, two main SDP scenarios are used in the literature: Within-Project Defect Prediction (WPDP) and Cross-Project Defect Prediction (CPDP). In WPDP, the historical data of a project (i.e., different versions) is used to predict the faulty parts (Omri & Sinz, 2020), i.e., WPDP focuses on fault predictions within the same software project on which it is trained (Ni et al., 2017). Hence, both the training and test sets belong to the same project. On the other hand, CPDP uses data from other projects (source projects) to train an SDP model and use this model to predict the faulty parts of another project (target project) (Chen et al., 2019). This approach originated from transfer learning and has particular importance where the target project may have inadequate labeled data for training. However, the main complication of this approach is to minimize the feature distribution difference between source projects and the target project.

The main obstacle in CPDP is that all projects used in the CPDP scenario must use the same metrics. In contrast, Heterogeneous Defect Prediction (HDP) enables defect prediction across projects with different metrics, mapping data from source and target projects into a common metric space (Chen et al., 2022; Nam et al., 2018).

In addition to these SDP scenarios, Just in Time Software Defect Prediction (JIT-SDP) is another popular approach (Kamei et al., 2013) aiming to predict software defects at the software change level (Cabral et al., 2019). JIT-SDP (also called change level defect prediction) enables developers identify and fix defects on time ensuring software quality. It has particular importance in SDP since it provides on time guidance to developers at a finer granularity, i.e., change level. Using JIT-SDP, developers can immediately review and test their changes without time consuming code reviews and extensive tests (Kamei et al., 2013; Zheng et al., 2022). A final novel SDP approach is Cross-Version Defect Prediction (CVDP) which utilizes the fault data in prior versions of the same project to predict the current version of the software project (Zhang et al., 2020b).

## 2.2 Deep Learning in SDP

Deep Learning is a subfield of ML based on Artificial Neural Networks (ANNs), simply called Neural Networks (NNs) with multiple layers (Arar & Ayan, 2015). The neural network model yields a prediction for each input in supervised, unsupervised, or semi-supervised training (Jorayeva et al., 2022). First, the error between these predictions and the actual results is calculated according to a previously simplified loss function. Later, the gradients of this function concerning the model's parameters are computed in a process called backpropagation and used for updating them in the next step with the help of an optimizer (Goodfellow et al., 2016). This section briefly introduces the most used DL models in SDP.

Multi-Layer Perceptron (MLP) is the fundamental structure of feedforward neural networks and has multiple layers (Goodfellow et al., 2016). The input layer contains the vectorized input data; hidden layers of interconnected nodes allow the structure to learn transformations on the data. The MLP model learns weights and biases using the backpropagation mechanism and nonlinear activation functions, extracting more advanced features. Finally, the output layer produces output vectors that correspond to the model's prediction of the input's class.

Deep Neural Networks (DNNs) are particular ANNs devised to learn by multi-connection layers (Montavon et al., 2018). The architecture of DNNs includes one input layer, one output layer, and one or more hidden layers between them. The input feature space of the data constitutes the input layer of the DNN. The input can be constructed with feature extraction methods. The output layer has one node in binary classification and has nodes as many as the number of classes in multi-class classification. DNN uses a standard backpropagation algorithm with a nonlinear activation function like sigmoid or Relu (Apicella et al., 2021). With the help of the defined architecture, DNNs extract features from the input data. Then, the model is trained to optimize weight and bias values in the neural network structure. Finally, the trained model is used to predict the class of the new input.

Inspired by the human visual system, Convolutional Neural Network (CNN) uses convolution operations to extract input features. This process is implemented through the multiple sliding kernels, matrices of specified shape and size, also called filters, and the elementwise multiplication of these kernels with the corresponding image data. This operation yields information about various features in the input. The CNN structure is very commonly used in image processing. For example, 1D convolution operations are applicable in diverse areas, including examining sequential data, text, or time-series data to find the patterns in the data (Rao & McMahan, 2019).

Recurrent Neural Networks (RNNs) have feedback loops in their architecture, allowing information to be memorized in short terms. Due to this property, RNN can analyze sequential input, such as speech, audio, weather, and financial data. However, an RNN's output at a stage relies on the previous output and the current input. While CNN shares unknown parameters across space, RNN shares them across time. Nevertheless, RNNs' memory is short-termed, their computation can be slow, and they suffer from the vanishing or exploding gradients problem (Hochreiter & Schmidhuber, 1997). Hochreiter & Schmidhuber (1997) developed the Long Short-Term Memory (LSTM) model to solve the problems with RNN structure. To overcome these issues, additional neural networks, called gates, are introduced, which handle the information stream in the network. Another type of RNN, Gated Recurrent Units (GRUs), were proposed by Chung et al. (2014). In GRUs, the gated approach solves RNN's information flow problems in long sequences with a simpler architecture introducing two gates: the update and the reset gates. Since GRUs have fewer gate structures than LSTMs', they have fewer parameters to change during the training, which leads them to being faster.

Deep Belief Neural Networks (DBNs) are feedforward Neural Networks (NNs) with many layers (Golovko et al., 2014). A DBN is not the same as the traditional types of DNNs discussed so far. Instead, a DBN is a particular DNN with undirected connections between some layers. These undirected layers are Restricted Boltzmann Machines and can be trained using unsupervised learning algorithms.

Encoder-Decoder models (also known as Sequence to Sequence or Seq2Sec models) are commonly used DNN architectures to convert input data in a particular domain into output data in another domain via a two-stage network (Cho et al., 2014; Chollampatt & Ng, 2018). First, the encoder takes a variable-length sequence in a specific domain and compresses it to a fixed-length representation. Then, the decoder maps the encoded data to a variable-length output in another domain. Due to these features, encoder-decoder models are widely used in many application areas, such as machine translation (Cho et al., 2014; Chollampatt & Ng, 2018).

Autoencoders might be considered as specific types of encoder-decoder models (Zhu et al., 2020). Autoencoder is an unsupervised ANN that learns efficient encoding of unlabeled data. First, it learns how to efficiently compress and encode input data. Next, autoencoders learn how to in the data and reduce data dimensions. Then, using the encoded representation, it learns how to reconstruct the data as close as to the original input. Autoencoders are used in many deep learning tasks such as anomaly and face detection. In addition, modified versions of autoencoders are used for specific tasks in deep learning. For example, sparse and denoising autoencoders are used in learning representations for subsequent classification tasks. Variational autoencoders are used in generative tasks to produce similar outputs to the input data. In SDP, autoencoders' main use is to extract features of input data automatically (Tong et al., 2018; Zhu et al., 2020; Wu et al., 2021; Zhang et al., 2021b).

Extreme learning machines (ELMs) are special feedforward neural networks invented by Huang et al. (2006). ELM architecture includes single or multiple layers of hidden nodes whose parameters need not be tuned. The hidden nodes of ELMs can be assigned randomly. No update operation is performed for these nodes, or they can be inherited from their antecessors without being changed. Generally, these nodes' weights are learned in a single step converging to a linear model. Hence, these models might be much faster than backpropagation-based neural networks. Moreover, these models might produce comparable results with SVM in classification and regression tasks (Liu et al., 2005; Liu et al., 2012).

Generative Adversarial Networks (GANs) are another approach used in generative modeling, designed by Goodfellow et al. (2014). Generative modeling is an unsupervised task. GANs use deep learning methods, such as CNN, to produce new outputs similar to the input data acquired from the original dataset. GANs use two neural networks named generator and discriminator. The generator is a CNN, and the discriminator is a de-convolutional NN. These networks compete in a game where one agent's gain is another agent's loss to predict more accurately. In this game, the generator produces artificial data similar to the real data, and the discriminator tries to distinguish the artificially generated data from the original data. The generator produces better artificial outputs as the game continues, and the discriminator will detect them better. In this way, GANs learn to generate new data with the same statistics as the training set.

Siamese Neural Networks (SNNs) are NNs that contain two or more subnetworks whose configurations, parameters, and weights are the same. Moreover, parameters are updated in both networks in the same way. In this way, Siamese NN compares its feature vectors and finds the similarity of the inputs by learning a similarity function. Hence it can be trained to check whether two inputs (for example, images of a person) are the same. Hence, SNN's architecture enables new data classification without retraining the network and making them suitable for one-shot learning problems. Furthermore, Siamese NNs are robust to class imbalance and learn semantic similarity. So, SNNs were used in several types of research in the SDP domain, although they require more training time than NNs (Zhao et al., 2018; Zhao et al., 2019).

Hierarchical Neural Network (HNN) is a special NN that consists of multiple loosely coupled subnets defined in the form of an acyclic graph. The subnets of the graph can be single neurons or complex NNs. Each subnet tries to acquire a specific figure of the input data (Mavrovouniotis & Chang, 1992). They are being used in various deep learning-based tasks such as classification (Wang et al., 2012) and image interpretation (Behnke, 2003). Further, HNNs have been used in SDP to provide better fault predictions (Wang et al., 2021; Yu et al., 2021a).

Graph Neural Networks (GNNs) are NNs designed to leverage the structure and properties of graphs. GNNs perform inference on data described by graphs by using deep learning methods. Hence, GNNs can be used in graph operations performing node-level, edge-level, and graph-level predictions. GNNs are active research topics in many domains such as social networks, knowledge graphs, and recommender systems (Chen et al., 2021; Kumar et al., 2022). GNNs are also used in the SDP domain to take full advantage of the tree structure of the source code. To this end, GNNs are exploited to acquire the inherent defect information of faulty subtrees, which are excluded based on a fix-inducing change (Xu et al., 2021a).

## 2.3 Related Work

Researchers use ML models obtained from SE data (source code, requirement specifications, test cases, etc.) to effectively and efficiently engineer software (Giray, 2021). Watson et al. (2020), Yang et al. (2020), and Ferreira et al. (2021) surveyed how DL has been used to solve SE problems in general. Yang et al. (2020) identified that SDP is the most popular sub-problem under testing and debugging problems for which DL is applied. Ferreira et al. (2021) found out that SDP is one of the top three problems SE researchers are dealing with. Although a few papers that use DL for SDP are included in these reviews, their number is very low compared to our study due to their search strings involving terms that are more generic (like "software engineering"). In addition, these reviews do not include a detailed analysis and synthesis of SDP studies.

About a decade ago, researchers started to synthesize the results of the studies to understand the progress in SDP. Catal & Diri (2009) analyzed 74 studies published between 1990 and 2007 to present a consolidated view of the use of ML and statistical techniques for SDP. They observed a significant increase in the number of primary studies in 2007 compared to previous years. Hall et al. (2011) synthesized the quantitative and qualitative results of 36 results published from January 2000 to December 2010. Malhotra (2015) analyzed 64 studies to understand the use of ML for SDP for the period of 1991 and 2013. In contrast to these studies, this study focuses on the synthesis of the studies that used DL algorithms for SDP.

Some review studies targeted a specific subarea of SDP. Hosseini et al. (2017) and Goel et al. (2017) focused on CPDP. Özakıncı and Tarhan (2018) reviewed the studies on early SDP, which utilized the metrics gathered earlier in the software development life cycle, such as metrics on requirements, design artifacts, and source code. Radjenović et al. (2013) identified software metrics and assessed their applicability in SDP. Li et al. (2020) investigated the use and performance of unsupervised learning techniques in SDP. Matloob et al. (2021) examined ensemble learning techniques for SDP. Different from these review studies, this study focuses on the use of DL for SDP in general.

Table 1 lists the related work on the use of DL for SDP. Eight of these studies did not follow a systematic review research method. Son et al. (2019) conducted a systematic mapping study on the use of ML and DL for SDP by examining 156 studies. Malhotra (2020) analyzed 20 primary studies to explore the use of DL for software quality prediction. Pandey et al. (2021) recently published a systematic review study on the use of ML and DL for SDP. Their study covers the primary studies published until June 2019 of which 36 are addressing DL for SDP.

Recently, Batool & Khan (2022) and Pachouly et al. (2022) published two SLR papers on the use of ML and DL for SDP. Both include studies published between 2010 and 2021 (both papers were submitted to a journal before the end of 2021). In addition, both studies included traditional ML and data mining techniques besides DL in their review. The SLR by Batool & Khan (2022) covers 11 primary studies focusing on DL. Pachouly et al. (2022) did not report the number of primary studies particularly using DL; they rather mention that they include 146 primary studies in total. On the other hand, their analysis covers only DBN, CNN, RNN/LSTM, and MLP excluding other types of DL approaches such as encoder-decoder architectures, GAN, and hybrid DL models. Compared to these two SLRs, this study reflects the state-of-the-art of the use of DL for SDP by analyzing 102 primary studies. Batool & Khan (2022) and Pachouly et al. (2022) provided information on the datasets, evaluation metrics, and ML/DL approaches used in SDP. Additionally, Pachouly et al. (2022) provided an analysis on the tools/frameworks and challenges related to datasets, such as class imbalance. This study includes analyses on these items except tools and frameworks and additionally presents information on the representation of source code, granularity level of prediction, validation approaches, and reproducibility package. Finally, this study includes a qualitative analysis on the challenges and proposed solutions on the whole aspects of SDP unlike the analysis of Pachouly et al. (2022) focusing on the challenges related to datasets.

This study differs from the related studies by involving a unique combination of the following characteristics: (1) focusing particularly on the use of DL for SDP, (2) with substantial level of depth on several aspects of DL-based SDP, (3) achieving a good coverage of the literature including 102 primary studies published until the end of 2021, and (4) following a systematic literature review research method.

**Table 1. Summary of related work**

| Reference | Year | Type of review | # of primary studies | Time period covered | Scope |
|---|---|---|---|---|---|
| Li et al. (2018) | 2018 | Non-systematic | 70 | Jan 2014 – Apr 2017 | ML & DL for SDP |
| Kalaivani & Beena (2018) | 2018 | Non-systematic | not reported | not reported | ML & DL for SDP |
| Prasad & Sasikala (2019) | 2019 | Non-systematic | not reported | not reported | ML & DL for SDP |
| Rathore & Kumar (2019) | 2019 | Non-systematic | not reported | 1993 – 2017 | ML & DL for SDP |
| Son et al. (2019) | 2019 | Systematic mapping | 156 | 1995 – 2018 | ML & DL for SDP |
| Omri & Sinz (2020) | 2020 | Non-systematic | not reported | not reported | ML & DL for SDP |
| Guan et al. (2020) | 2020 | Non-systematic | not reported | not reported | ML & DL for SDP |
| Malhotra et al. (2020) | 2020 | Systematic mapping | 20 | Jan 1990 – Jan 2019 | DL for software quality prediction |
| Akimova et al. (2021) | 2021 | Non-systematic | not reported | 2019 – 2021 | DL for SDP |
| Atif et al. (2021) | 2021 | Non-systematic | not reported | not reported | Statistics, ML & DL for SDP |

| Pandey et al. (2021) | 2021 | Systematic literature review | 154 out of which 36 are on DL | 1990 – June 2019 | ML & DL for SDP |
| Batool & Khan (2022) | 2022 | Systematic literature review | 68 out of which 11 involve DL | 2010 - 2021 (partial) | Data mining, ML & DL for SDP |
| Pachouly et al. (2022) | 2022 | Systematic literature review | 146 (number of studies involving DL not reported) | 2010 - 2021 (partial) | ML & DL |
| This study | 2022 | Systematic literature review | 102 | until the end of 2021 | DL for SDP |

## 3 RESEARCH OBJECTIVES AND METHOD

This section describes the research objectives and the method used in this study. We adopted a systematic literature review (SLR) approach to synthesize the knowledge on the use of DL algorithms for SDP. The research method is based on established guidelines (Kitchenham and Charters, 2007; Wohlin, 2014), some previous good examples of SLRs on SDP (Hall et al., 2011; Hosseini et al., 2017), and our previous experience in conducting SLRs (Garousi et al., 2019; Giray and Tüzün, 2018; Tarhan and Giray, 2017). Table 2 summarizes the SLR protocol used in this study using the format adopted from Motta et al. (2018). We performed four main activities: (1) defining the goal and the research questions (RQs), (2) selecting relevant primary studies, (3) extracting data, and (4) synthesizing data and reporting the results. The details are described in the following subsections.

**Table 2. Protocol summary**

| Themes addressed by RQs | **RQ1.** SDP scenarios (e.g., WPDP, CPDP, HDP) |
| --- | --- |
| | **RQ2.** ML categories (i.e., supervised/unsupervised/semi-supervised learning) |
| | **RQ3.** Training and testing datasets |
| | **RQ4.** Representation of source code (e.g., metrics, Abstract Syntax Tree) |
| | **RQ5.** Granularity level of prediction (e.g., file, change, class) |
| | **RQ6.** Techniques for dealing with the class imbalance problem (e.g., over-sampling, under-sampling) |
| | **RQ7.** DL approaches (e.g., CNN, LSTM) |
| | **RQ8.** Evaluation metrics (e.g., F-measure, recall, AUC; etc.) and validation approaches (i.e., cross-validation and hold-out) |
| | **RQ9.** Reproducibility package |
| | **RQ10.** Challenges and proposed solutions |
| Search string | Population: software defect/fault/bug/quality prediction/estimation |
| | Intervention: deep learning |
| | *(software) AND (fault OR defect OR quality OR bug) AND (predict\* OR estimat\*) AND ("deep learning")* |
| Search strategy | DB search: ACM, IEEE, ScienceDirect, Springer, Wiley |
| | Manual search on the primary studies included in the secondary studies listed in Table 1 |
| | Forward snowballing using Google Scholar |
| Inclusion and exclusion criteria | Inclusion: |

- The paper is written in English
- The paper is published in a scholarly journal or conference/workshop/symposium proceedings.
- The paper involves at least one DL algorithm applied to SDP problem and reported empirical results.

Exclusion:

- The paper's full text is not available.
- The paper is an editorial, issue introduction or secondary study (literature review, SMS, SLR).
- The paper involves only traditional ML algorithms or statistical techniques applied to SDP problem.

| | |
|---|---|
| Study type | Primary studies |

## 3.1 Goal and Research Questions

The scope and goal of this study were formulated using the Goal-Question-Metric approach (Basili et al., 1994) as follows.

***Analyze*** *the state-of-the-art in software defect prediction (SDP)*

***for the purpose of*** *exploration and analysis*

***with respect to*** *the SDP scenarios, ML categories, datasets, representation of source code, granularity of prediction, techniques for dealing with the class imbalance problem, DL algorithms, evaluation metrics and validation approaches, presence of a reproducibility package, and reported challenges and proposed solutions*

***from the point of view of*** *machine learning researchers*

***in the context of*** *deep learning (DL).*

The goal of this study is to systematically classify, review, and synthesize the body of knowledge and evidence on the use of DL algorithms for SDP. As Kitchenham et al. (2015) pointed out, RQs must embody secondary studies' goals. Accordingly, we raised the following RQs to achieve our goal:

**RQ1.** Which SDP scenarios (e.g., WPDP, CPDP or HDP) were applied?

**RQ2.** Which ML categories (i.e., supervised/unsupervised/semi-supervised learning) were applied in DL-based SDP studies?

**RQ3.** Which public datasets were used for the development and testing of ML/DL models for SDP?

**RQ4.** How did researchers represent source code to develop DL models for SDP?

**RQ5.** At which granularity levels did researchers perform SDP?

**RQ6.** What approaches did researchers follow to cope with class imbalance challenge for SDP?

**RQ7.** Which DL algorithms (e.g., CNN, LSTM) were applied?

**RQ8.** What kind of evaluation metrics and validation approaches were used?

**RQ9.** How often did researchers provide reproduction packages to support the reproducibility of DL models for SDP?

**RQ10.** What were the challenges and proposed solutions in the use of DL for SDP?

Figure 2 shows how the RQs are mapped to the ML model life cycle proposed by Amershi et al. (2019). The feedback loops and the iterations throughout the life cycle were omitted for the sake of simplicity. In the model requirements stage, researchers decide on the SDP scenario(s) they will focus and select the most appropriate type of ML category for the SDP problem. During the data engineering stages, teams look for available datasets or construct new ones, clean data if required, and prepare labeled datasets for supervised learning if labels are not already present. Two basic decisions affect the data engineering phase, i.e., how to represent source code (such as via metrics, abstract syntax trees) and the granularity level of prediction (such as file-level, function-level). A very common problem researchers deal with in data engineering phase is class imbalance. Software defect datasets generally have fewer buggy modules than non-buggy ones (Bennin et al., 2017). Unfortunately, such datasets consisting of imbalanced data typically decrease the prediction performance of ML/DL models developed for SDP (Bennin et al., 2017). Feature engineering refers to the activities for extracting and selecting informative features for ML models (Amershi et al., 2019). For some approaches using DL algorithms, the feature engineering stage is less explicit and combined with the model training stage (Amershi et al., 2019). Generally, researchers use more than one ML/DL algorithm to develop models for SDP. In model evaluation, teams evaluate output models using evaluation metrics and approaches to choose the best-performing model. To enable other researchers to obtain the reported experimental results with the same experimental setup, researchers must share source code and datasets (Liu et al., 2021). During this model life cycle, teams may face some challenges. After model development, the chosen model is deployed and monitored in a production environment. We excluded these two stages since our primary studies did not include any information on these stages.

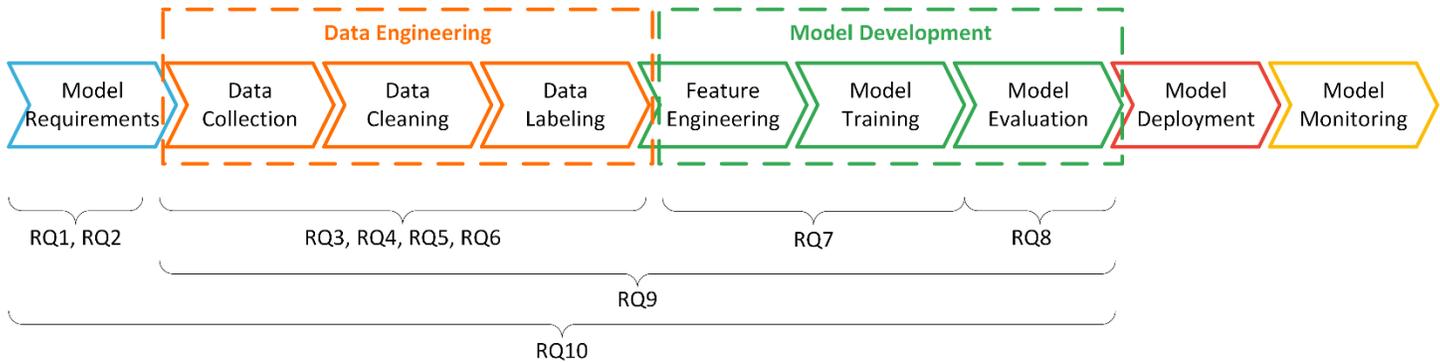

**Figure 2. RQs mapped to the ML model life cycle adapted from Amershi et al. (2019)**

## 3.2 PRIMARY STUDY SELECTION

Figure 3 depicts the process we used for primary study selection. In the first step, we used five widely used online databases, i.e., ACM, IEEE Xplore, ScienceDirect, Springer, and Wiley as the source of the potentially relevant primary studies. We used the following search string to query these databases (for details see Appendix 8.1): *(software) AND (fault OR defect OR quality OR bug) AND (predict\* OR estimat\*) AND ("deep learning")*. We used paper title, abstract, and keywords as the search fields. We searched each of the five online databases two times in June 2021 and January 2022. In January 2022, we searched the databases to obtain the papers published only in the second half of 2021 and added the new candidate papers to our pool. By doing so, we aimed to involve all primary studies published until the end of 2021. As shown in Table 3, we obtained 296 primary studies in total (238 in June 2021 and additional 58 papers in January 2022) to apply inclusion and exclusion criteria.

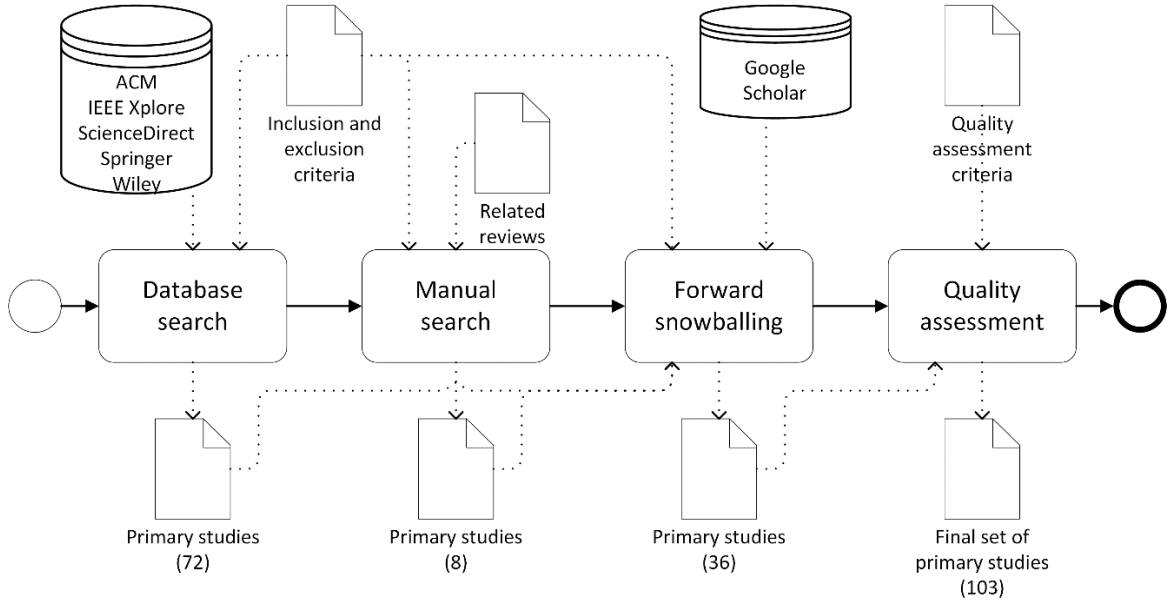

**Figure 3. The primary study selection process**

**Table 3. Database search results**

| Database | Search in June 2021 | Search in January 2022 |
|---|---|---|
| ACM | 22 | 3 |
| IEEE Xplore | 89 | 23 |
| ScienceDirect | 29 | 6 |
| Springer | 83 | 24 |
| Wiley | 22 | 2 |
| Duplicates removed | 7 | 0 |
| Total for inclusion/exclusion | 238 | 58 |

To identify the relevant set of papers to answer our RQs, we specified our inclusion and exclusion criteria, as presented in Table 2. Each primary study was assigned to two authors for the application of inclusion and exclusion criteria. Initially, two authors voted on the candidate papers individually whether to include by reading the title, abstract, keywords, and by checking the full text if needed. When the voting results were compared, there was 86% agreement between two authors. In the case of disagreements, two other authors were assigned to investigate whether to include a paper and the conflict was resolved. As the result of the first step, we obtained 72 primary studies for quality assessment.

In the second step, we checked the primary studies used in the related reviews listed in Table 1 to enrich our paper pool. We applied our inclusion and exclusion criteria to all studies that used DL for SDP in these reviews. This manual search led to the addition of eight primary studies to our pool.

In the third step, we conducted forward snowballing, as recommended by systematic review guidelines (Wohlin, 2014), to ensure the inclusion of as many relevant primary studies as possible. We opted for forward snowballing rather than backward snowballing since their efficiencies are similar, with forward snowballing finding slightly lower number of non-relevant papers (Badampudi et al., 2015), and not both to manage the manual effort needed. We checked the citations listed on Google Scholar to each primary study found in the first two steps, i.e., database and manual search, against the inclusion and exclusion criteria. Forward snowballing provided additional 36 primary studies.

In the last step, the authors conducted a quality assessment for the primary studies assigned to them before extracting data, as proposed in the literature (Hassler et al., 2014). Each primary study was assessed by one of the authors. Table 4 lists the criteria used for quality assessment. We derived these criteria from Kitchenham et al. (2009) and our earlier SLRs, such as Catal et al. (2021). We scored each paper using a 3-point Likert scale (yes = 1, somewhat = 0.5, no = 0) for each criterion. For instance, we scored for Q1 as 1 if the aim of the study was stated clearly in the introduction (expected place); as 0.5 if the aim was vaguely stated, or not at the expected place, and as 0 if the aim was not stated in the paper. We decided to include the papers with a score higher than four points to maintain a high-quality input of primary studies. We excluded 14 studies (listed in Appendix 8.2) with a score under our threshold. As the result of the primary study selection step, we obtained a total number of 102 papers for data extraction.

**Table 4. Quality assessment criteria**

| # | Question |
|---|---|
| Q1 | Are the aims of the study clearly stated? |
| Q2 | Are the scope and context and experimental design of the study clearly defined? |
| Q3 | Are the variables in the study likely to be valid and reliable? |
| Q4 | Is the research process documented adequately? |
| Q5 | Are all the study questions answered? |
| Q6 | Are the negative findings presented? |
| Q7 | Are the main findings stated clearly (regarding creditability, validity, and reliability)? |
| Q8 | Do the conclusions relate to the aim of the purpose of the study, and are they reliable? |

### 3.3 DATA EXTRACTION

After primary study selection, we started with the data extraction phase. We formed an initial data extraction form (Table 5) based on our RQs. The first six rows constitute the metadata of the papers. The first author formed an initial list of categories using previous SLRs (Catal et al., 2021) and conducted a pilot data extraction on a few randomly selected primary studies. Afterwards each author extracted data from the primary studies assigned to him. Whenever an author was undecided about the data to be extracted, he recorded that case, and these cases were resolved via discussions among the authors. During data extraction phase, we continuously refined the categories iteratively and incrementally during data extraction. We recorded the reported challenges and proposed solutions as free text for further analysis and synthesis.

**Table 5. Data extraction form**

| Field | Input Type/Categories | Relevant RQ |
|---|---|---|
| Paper ID | Auto incremented number | - |
| Paper title | Free text | - |
| Abstract | Free text | - |
| Keywords | Free text | - |
| Publication year | Number | Demographics |
| Venue/Journal/Conference | Free text | Demographics |
| SDP scenario | Multiple selection | RQ1 |
| ML category | Multiple selection | RQ2 |
| Dataset | Multiple selection | RQ3 |
| Representation of source code | Multiple selection | RQ4 |
| Granularity level of prediction | Multiple selection | RQ5 |
| Dealing with class imbalance problem | Multiple selection | RQ6 |
| DL algorithms used | Multiple selection | RQ7 |
| Evaluation metrics | Multiple selection | RQ8 |
| Validation approach | Multiple selection | RQ8 |
| Reproducibility package | Multiple selection | RQ9 |
| Challenges and proposed solutions | Free text | RQ10 |

### 3.4 Data Synthesis and Reporting

We extracted quantitative data using categories for the RQs between one and nine. Thus, we reported the frequencies and percentages of each identified category to answer these RQs.

The only RQ that required qualitative analysis was RQ10, i.e., the challenges and proposed solutions. 50 out of 102 primary studies reported one or more challenges and some of them proposed solutions to cope with these challenges. We recorded the challenges and the proposed solutions in Google sheets during data extraction. We conducted open coding (Miles et al., 2019) to analyze the challenges. A code symbolically assigns a summative or evocative attribute for a portion of qualitative data (Miles et al., 2019). We performed open coding in cycles. In the first cycle, we identified any emerging patterns of similarity or contradiction. In the second cycle, we collapsed and expanded to understand any patterns. After extracting the main themes and codes, we revised the codes and assigned them to each challenge.

### 4 Results

This section presents the responses to the RQs defined at the beginning of this research study. Before presenting the responses, we provide additional information about the identified primary studies, e.g., the yearly distribution and distribution of the studies per venue. As Figure 4 shows, there is an increasing trend in the number of primary studies. This indicates that the application of DL algorithms for SDP is a recent trend among researchers, especially ongoing as of 2019.

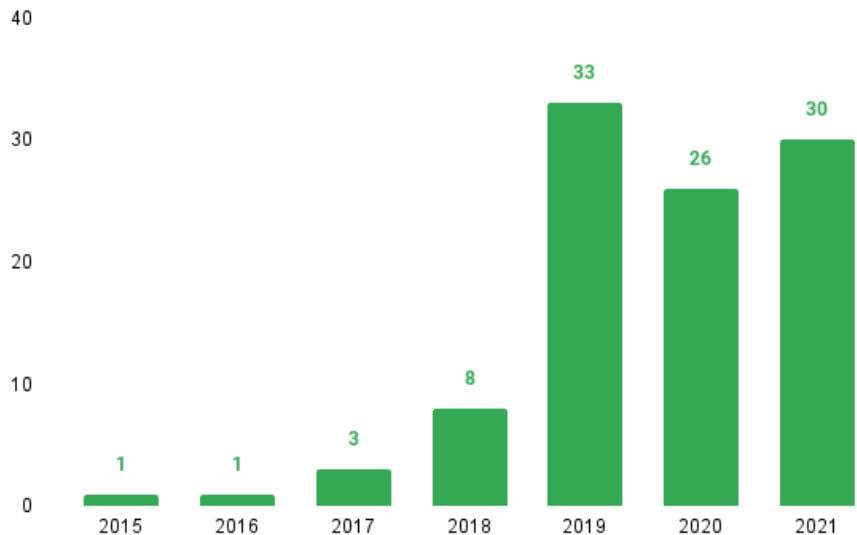

**Figure 4. Number of primary studies over the years**

54% of the studies (55 primary studies) were published in journals. The journals with the most papers are IEEE Access (10 papers) and IET Software (five papers). 46% of the studies (47 primary studies) were presented in conferences and workshops. The conference with the most papers (four papers) is the International Conference on Software Quality, Reliability and Security (QRS). Appendix 8.3 includes the list of the venues in which the primary studies were presented and published.

## 4.1 SDP Scenarios

Figure 5 shows the number of studies per SDP scenario. While 82 studies include WPDP, 42 studies involve CPDP. 22 studies encompass experiments for both WPDP and CPDP.

Ten studies in total focus on cross-version defect prediction. Six of these are classified under WPDP scenario. In one of these studies, Zhang et al. (2021a) conducted experiments on total 32 cross-version pairs derived from 45 versions of 13 software projects obtained from PROMISE (Jureczko & Madeyski, 2010), NASA (Shepperd et al., 2013), and SOFTLAB (Turhan et al., 2009) repositories. For instance, in three of the experiments, they trained a model using Ant versions 1.3, 1.4, and 1.5 and tried to predict bugs in Ant versions 1.4, 1.5, and 1.6, respectively (Zhang et al., 2021a). Li et al. (2019c) examined the ability of a model trained on all the existing versions of a project X and other projects to detect bugs on an unseen version of the project X, i.e., CPDP. Shi, et al. (2021) conducted experiments on cross-version defect prediction in both WPDP and CPDP settings. For instance, they built a model using Camel 1.4 and tried to predict bugs for Camel 1.6 (WPDP) and Jedit 4.1 (CPDP).

Young et al. (2018) conducted experiments on just-in-time defect prediction, a.k.a. change level defect prediction, using six open-source projects. For each project, they built models via a training set obtained from that project and tried to predict defect-prone changes for the same project. Xu et al. (2021b) emphasized the difficulty of collecting sufficient labeled bug data for some mobile applications. Hence, they proposed to learn a high-level feature representation from a bug dataset consisting of 19 mobile applications for JIT defect prediction. Zeng et al. (2021) used DL to build models for identifying defective commits in both WPDP and CPDP settings.

Three studies addressed HDP. Gong et al. (2019) designed a neural network to deal with heterogeneous metric sets for defect prediction. Sun et al. (2021) proposed a deep adversarial learning based HDP method. Wu et al. (2021) proposed a method for multi-source heterogeneous cross-project defect prediction. They used an autoencoder to extract the intermediate features from the original datasets instead of simply removing redundant and unrelated features (Wu et al., 2021).

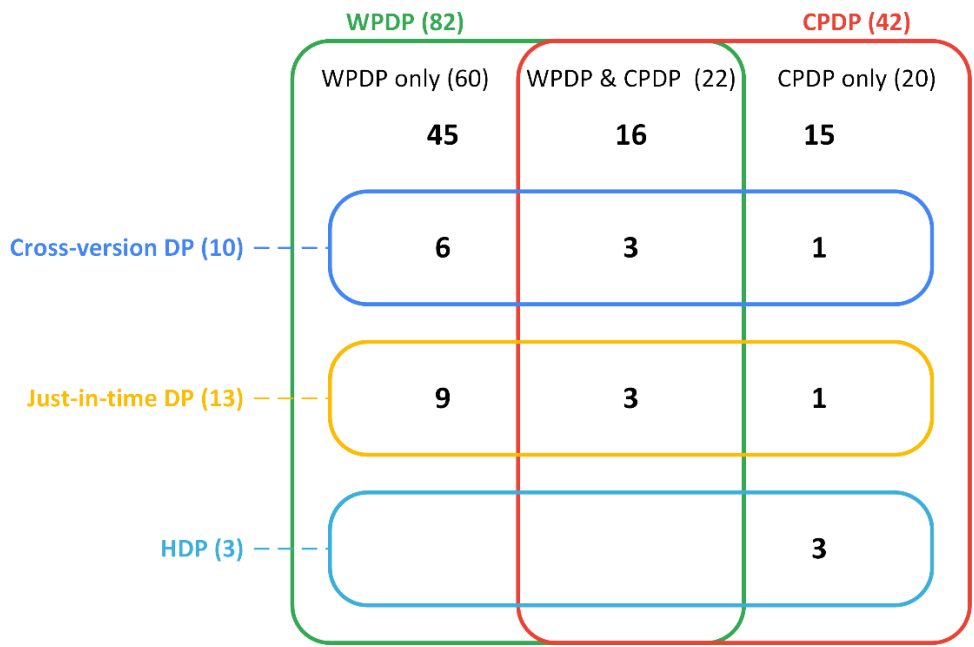

**Figure 5. Number of primary studies per SDP scenarios**

Researchers started to show interest in cross-version defect prediction using DL in last three years. Our paper pool includes two studies published in 2019 and four in 2020 and 2021. The first study on JIT defect prediction with DL dates back 2015. One study published in 2018 and the rest 11 studies were published in last three years. One study on HDP was published in 2019 and the other two in 2021.

## 4.2 ML Categories

The second RQ is related to the ML categories (i.e., supervised learning, unsupervised learning, and semi-supervised learning). 94% of the primary studies (96 studies) apply supervised DL. 77 of these do not include any other ML category. 23 studies involve unsupervised DL learning. While three of them do not include any other ML category, 19 of them involve supervised and one of them involves semi-supervised learning.

Shi et al. (2021) conducted experiments using unsupervised and semi-supervised learning. Two studies, i.e., Sun et al. (2020a) and Xu et al. (2021b) involve only semi-supervised learning.

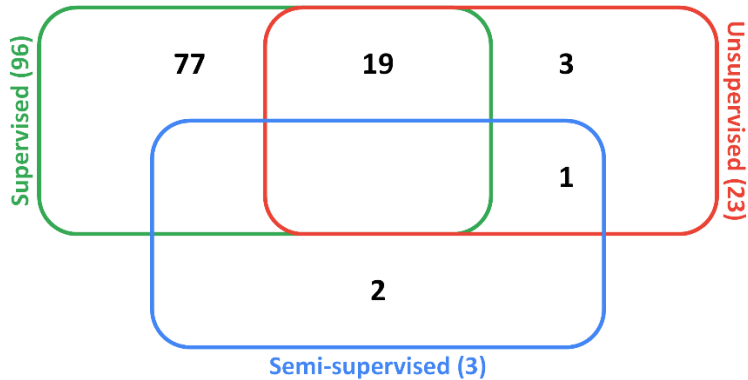

**Figure 6. Number of primary studies per ML category**

All the studies, except one (Wang et al., 2016), published before 2018 includes only supervised learning. The first three studies involving unsupervised learning were published in 2018 (Bhandari & Gupta, 2018; Sun et al., 2018; Tong et al., 20187). 10 of 23 studies including unsupervised learning were published in 2019. Afterwards, we see a downward trend in the use of unsupervised learning, four study in 2020 and five in 2021. Semi-supervised learning for SDP with DL started to be used after 2020 (Sun et al., 2020a; Xu et al., 2021b; Shi et al., 2021).

## 4.3 Training and Testing Datasets

To answer this RQ, we extracted the datasets used in each primary study. Researchers used more than eight datasets on average per study. Table 6 lists the datasets that were used in ten or more primary studies. All these datasets were developed in one of three programming languages, i.e., Java, C, and C++. The top 11 most frequently used datasets are from PROMISE (Jureczko & Madeyski, 2010) repository and were developed using Java.

**Table 6. The datasets used in ten or more primary studies**

| Project | # of Primary Studies | Programming Language | Description |
|---|---|---|---|
| Camel | 55 | Java | Enterprise integration framework |
| Xalan | 53 | Java | A library for transforming XML files |
| Xerces | 49 | Java | XML parser |
| Poi | 48 | Java | Java library to access Microsoft format files |
| Log4j | 44 | Java | Logging library for Java |
| Lucene | 44 | Java | Text search engine library |

| Synapse | 44 | Java | Data transport adapters |
|---|---|---|---|
| Jedit | 43 | Java | Text editor designed for programmers |
| Ant | 39 | Java | Java based build tool |
| Ivy | 32 | Java | Dependency management library |
| Velocity | 22 | Java | A Java-based template engine |
| Eclipse JDT | 18 | Java | Eclipse Java Development Tools |
| PC1 | 18 | C | A flight software for earth orbiting satellite |
| CM1 | 16 | C | A NASA spacecraft instrument |
| KC1 | 16 | C++ | A system implementing storage management for receiving and processing ground data |
| PC3 | 16 | C | NASA orbiting project |
| PC4 | 16 | C | NASA satellites project |
| MW1 | 15 | C | Zero gravity experiment |
| JM1 | 14 | C | A real-time predictive ground system |
| KC2 | 13 | C++ | Data from C++ functions. Science data processing; another part of the same project as KC1; different personnel than KC1.  Shared some third-party software libraries with KC1, but no other software overlap. |
| MC1 | 13 | C & C++ | NASA combustion project |
| PC2 | 11 | C | NASA for earth project |
| MC2 | 10 | C | One of the NASA Metrics Data Program defect data sets |

Some studies, e.g., Zeng et al. (2021), conducted experiments on the projects developed via C++ (QT and OpenStack) and Java (JDT, Platform, and Gerrit) as well as the projects like Go developed with a modern programming language, like Golang. Dong et al. (2018) focused on predicting bugs in Android binary executables called "apk"s. They obtained Android projects, such as Wikipedia and Chess apps, from GitHub and constructed datasets to build models for defect prediction. Xu et al. (2021b), Zhao et al. (2021a), and Zhao et al. (2021b) also used Android projects to build and test bug prediction models.

Ferenc et al. (2020) used a public unified bug dataset for Java (Ferenc et al., 2018), which is an amalgamation of three repositories, i.e., PROMISE (Jureczko & Madeyski, 2010), the Bug Prediction Dataset (D'Ambros et al., 2010), and the GitHub Bug Dataset (Tóth et al., 2016). Xu et al. (2021a) and Zhang et al. (2020a) crawled GitHub and Codeforces, respectively to build datasets. Only one study, i.e., Albahli (2019), included commercial projects in their experiments. They tried to build models and predict bugs using six open source and five commercial projects (Albahli, 2019).

### 4.4 REPRESENTATION OF SOURCE CODE

While source code comprises of textual data, DL algorithms work on numeric vectors (Gousios, 2021). Source code representation refers to converting source code to a form that can be processed by DL algorithms. This conversion process should consider and optimize the loss of information during conversion. Figure 7 shows source code representation approaches used in the primary studies in our pool. Different kinds of metrics, i.e., software size and structure, process, and product metrics, are the most frequently used representation technique for SDP. 40 studies used an intermediary representation (some of them involve more than one), i.e., Abstract Syntax Tree (AST), assembly code, Control Flow Graph, Program Dependency Graph, Data Flow Graph, and image, to form a numeric vector to be fed to a DL algorithm. Eight studies combined metrics with an intermediary representation. Two studies converted source code directly to a numeric vector.

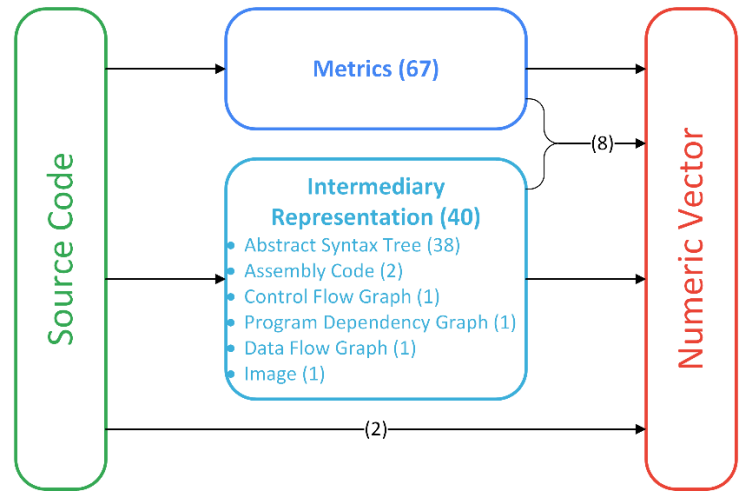

**Figure 7. Source code representation approaches used in the primary studies**

67 studies used a set of metrics to represent source code. Some of the studies (such as Xu et al., 2019) used a tool like CKJM (Spinellis, 2005) to extract size and structure software metrics by processing the bytecode of compiled Java files. These metrics include weighted methods per class, coupling between object classes and McCabe's cyclomatic complexity. Process features or change features obtained from the change history of a software project are also an indicator for defect prediction (Rahman & Devanbu, 2013). For instance, Yang et al. (2015) utilized change metrics, such as number of modified files, number of developers that changed the modified files, lines of codes added and deleted for JIT defect prediction. Ardimento et al. (2021) used some metrics, such as commit frequency, developer seniority, owned commit, mean time between commits, to represent the development process. Product metrics, describing the source code internal structure quality, are another type of metrics used for defect prediction. Such metrics include number of attributes inherited, depth of inheritance tree, number of methods, and number of static methods. Ardimento et al. (2021) used product metrics along with process metrics. Some researchers, such as Tong et al. (2018) and Zhao et al. (2019), normalized the values of metrics before forming a numeric vector.

Abstract Syntax Tree (AST) is a tree representation of the abstract syntactic structure of source code (Mou et al., 2016). 38 studies involved AST as an intermediary representation to construct a numeric vector. Liang et al. (2019) converted source code to AST and extracted tokens from AST nodes to generate token sequences. These sequences were mapped to fixed-length vectors and a Continuous Bag of Words (CBOW) model was built using all datasets to be fed to an LSTM network. Chen et al. (2019) used a simplified version of AST, by considering node types, which are project-independent, and ignoring method and variable names, which are project-specific. Some researchers, such as Li et al. (2017), Dam et al. (2019), and Liu et al. (2020), used word embeddings to obtain numeric vectors from ASTs. Shi et al. (2020) built embedding vectors using an AST path pair-based source code representation method named PathPair2Vec.

Li et al. (2019b) modelled and analyzed the relations among paths of ASTs from different methods using Program Dependency Graph (PDG) and Data Flow Graph (DFG). While the local context of buggy code is represented by buggy paths in AST, the global context of buggy code is represented by the relations among buggy methods modelled via program and data flow dependencies (Li et al., 2019b).

Phan & Le Nguyen (2017) preferred assembly instruction sequences over ASTs since they may simulate program behaviour better due to its closeness to machine code and reflect program structure. Phan et al. (2017) leveraged control flow graphs constructed from assembly instructions obtained by compiling source code.

Chen et al. (2020) proposed source code visualization for SDP, in other words, they represented source code as images and trained image classification models that predict defects. Each source file was converted into a vector of 8-bit unsigned integers corresponding to the ASCII decimal values of the characters in the source code. Then, an image is generated from that vector to be fed to ImageNet's pre-trained AlexNet model for classification (Chen et al., 2020).

Eight of the studies combined an AST-based input with a set of metrics. Fan et al. (2019b), Li et al. (2017), Lin & Lu (2021), Qiu et al. (2019b), Shi et al. (2021) and Wang et al. (2021) combined word embeddings of ASTs with metrics. Fiore et al.

(2021) obtained a vector from the nodes of AST. Afterwards, they combined these vectors with the vectors involving metrics. Huo et al. (2018) extracted five types of metrics (authorship, change type, change interval, code churn, and co-change) by analyzing change logs and textual contents generated by version control systems. In addition, they analyzed the differences between ASTs and identified change semantic types, like insertion of an expression statement or change of an infix expression. Afterwards, they built change sequences out of these metrics and semantic information of changes and used them for training DL models. For instance, a sequence for authorship can be represented as <developer1, developer2, developer1, developer3, developer2>.

Two studies did not use any intermediary representation and converted source code directly into numeric vector. Hoang et al. (2019) parsed commit messages and code changes using NLTK (Loper & Bird, 2002) and represented each word in the commit messages and code changes as a n-dimensional vector. Tian & Tian (2020) converted source code into fixed length vectors using Word2vec (Mikolov et al., 2013).

### 4.5 GRANULARITY LEVEL OF PREDICTION

ML/DL models were constructed to predict defects at various levels of granularity (Nam, 2014), i.e., file, module, change, class, function, procedure, and statement. Previous research found out that level of granularity has an impact both on model prediction performance and effort required to localize defects (Koru & Liu, 2005; Calikli et al., 2009). 37 studies involved a model predicting defects at file level and 32 studies at module level. 12 studies built models to identify buggy changes. Four studies included class-level predictions for a software system developed using an object-oriented programming language. More fine-grained levels of granularity level predictions, i.e., function/procedure and statement/line level, were addressed by three primary studies each. 16 studies did not report any granularity level of prediction.

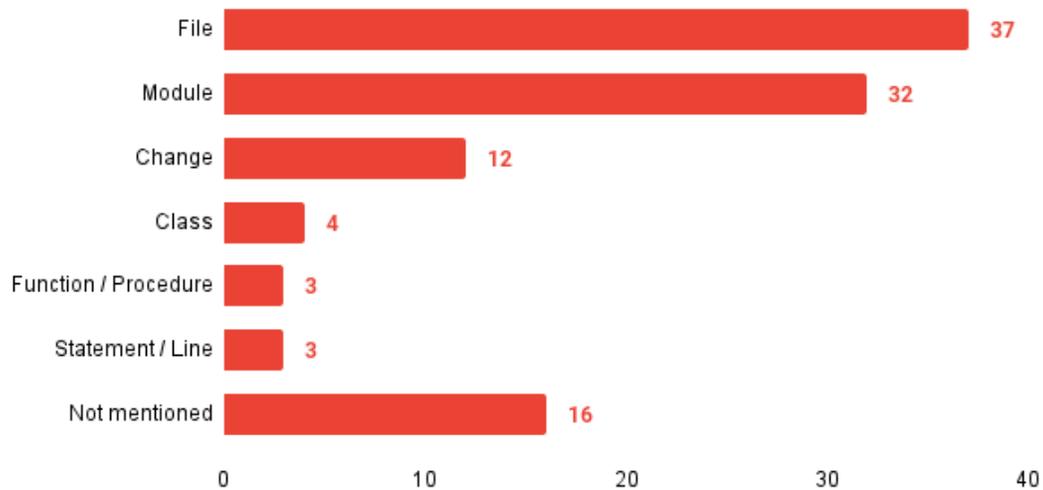

**Figure 8. Granularity levels of prediction reported in the primary studies**

In line with the observation of Kamei & Shihab (2016), researchers started to show more interest in using DL for SDP at more fine-grained levels. Since 2019, nine studies (two in 2019, four in 2020, three in 2021) reported the results of their experiments on class, statement, and procedure-level predictions. In addition, starting with one study in 2018, researchers published papers on change-level defect prediction using DL algorithms. 11 studies followed these in the last three years, i.e., three papers in 2020 and four papers in 2019 and 2021.

Three studies (Zhou et al., 2019; Wang et al., 2020; Zhu et al., 2021a) reported experiment results at more than one granularity level. Wang et al. (2020) deployed a DBN to learn semantic features from ASTs for file-level defect prediction models and source code changes for change-level defect prediction models automatically. DBN-based semantic features helped in improving prediction performance by varying percentages (from 2.9% to 13.3%) at file and change levels in WPDP and CPDP scenarios (Wang et al., 2020). Zhou et al. (2019) used different datasets to train and test DL models for file-, module-, and class-level defect prediction. Zhu et al. (2021a) used five datasets in their experiments, i.e., PROMISE (Jureczko & Madeyski, 2010) and AEEEM (D'Ambros et al., 2012) for class-level, NASA (Shepperd et al., 2013) and SOFTLAB (Turhan et al., 2009) at function-level, and ReLink (Wu et al., 2011) at file-level defect prediction.

### 4.6 TECHNIQUES FOR DEALING WITH THE CLASS IMBALANCE PROBLEM

Class imbalance problem arises when there is a severe skew in the class distribution in a dataset (Brownlee, 2020b). In SDP, datasets generally involve much fewer defective samples compared to non-defective ones. Having this bias, i.e., defective samples as the minority class, influences the prediction performance of DL models, sometimes leading to ignore the minority class entirely. This is a serious problem since it is important to predict defective instances. There are techniques to address class imbalance problem at data-level and algorithm level (Tong et al., 2019).

Figure 9 shows the frequencies of the techniques used by researchers to address class imbalance problem. Since six studies involved more than one technique, the total number of data points in the figure is more than 102. 63 of the studies used a technique at the data-level. They applied a kind of over-sampling or under-sampling or tried to create data to balance minority and majority classes. Nine studies tried to cope with the imbalanced datasets at the algorithm-level using either cost-sensitive or ensemble learning techniques. 42 of the studies did not report any technique to address class imbalance problem.

The most frequently used technique used at the data-level is oversampling (42 studies). Oversampling techniques duplicate or create new synthetic instances in the minority class (Brownlee, 2020b), i.e., creating new defective samples. 21 studies used Synthetic Minority Oversampling TEchnique, or SMOTE for short (Chawla et al., 2002) for oversampling. Some studies (Eivazpour & Keyvanpour, 2019; Xu et al., 2019) used ADAptive SYNthetic (ADASYN) sampling, which is an extension of SMOTE. 20 studies applied a kind of random oversampling. One study, i.e., Yedida & Menzies (2021), proposed a fuzzy sampling technique, which is a variation of oversampling. Eight studies used other techniques to create new instances of defective samples. Xu et al. (2021a) identified 3,026 bug fixes in 307 Java projects on GitHub. They constructed a dataset by combining the defective and fixed versions of the source files of these 3,026 bug fixes and ended up with a balanced dataset. Bhandari & Gupta (2020) increased the number of defective instances by injecting defects into the source files.

Zhang et al. (2021a) leveraged WGAN-GP (Wasserstein GAN with Gradient Penalty) to generate more defective training instances. Similarly, Sun et al. (2018), Eivazpour & Keyvanpour (2019) utilized Variational Autoencoder (VAE) and Zhu et al. (2021a) used GAN (Generative Adversarial Networks) to generate defective instances. Sun et al. (2020b) used both VAE and GAN along with SMOTE in their experiments.

Nine studies included algorithm-level techniques to address class imbalance problem. Six of them utilized cost-sensitive techniques. Li et al. (2019a) and Xu et al. (2019) proposed to assign different misclassification costs to the different classes in the model building stage to learn defective instances better. Similarly, Hoang et al. (2019) imposed a higher cost on misclassification of minority class, i.e., buggy commits, than they do on misclassification of the majority class, i.e., non-buggy commits to increase the performance of their JIT defect prediction model. Zhao et al. (2019) introduced a cost-sensitive cross-entropy loss function into DNN for JIT defect prediction in Android applications. Thus, they considered the prior probability of minority and majority classes, i.e., defective and clean commits respectively, when calculating cross-entropy loss to compensate class imbalance. Gong et al. (2019) assigned different misclassification costs to defective and non-defective instances to increase the performance of the neural network for HDP.

The authors of the three of the nine studies preferred an ensemble learning technique. Tran et al. (2019) used a two-stage ensemble learning method in the training stage. Tong et al. (2018) also reported that they used two-stage ensemble learning to cope with imbalanced datasets as well as eliminating overfitting problem. Xu et al. (2019) included Bagging, Balanced Bagging, AdaBoost, RUS with AdaBoost, EasyEnsemble, and BalanceCascade in their experiments.

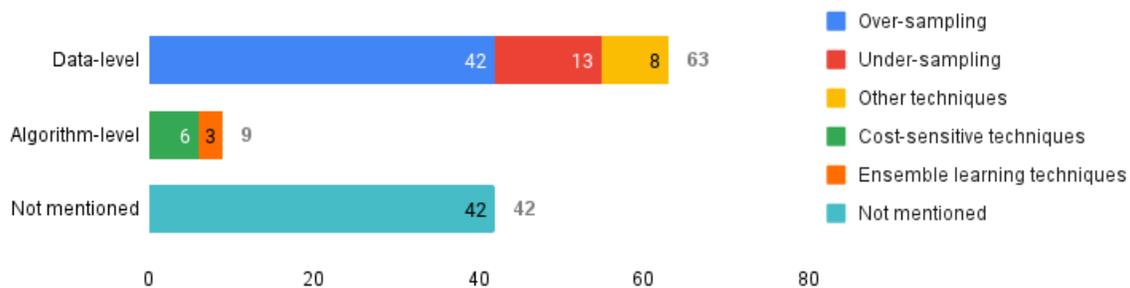

**Figure 9. Distribution of the techniques used for coping with class imbalance problem**

Six studies out of 102 used more than one technique to cope with class imbalance problem and reported their findings. Yedida & Menzies (2021) found out that oversampling is effective and necessary to applying DL for SDP. They applied oversampling, SMOTE, and fuzzy sampling on 24 datasets. Based on the F1 scores obtained from 240 data points (10 repeats over the 24 datasets), most significant improvements to DL model performance came from fuzzy sampling approach. Eivazpour & Keyvanpour (2019) applied various oversampling techniques, such as SMOTE, ADASYN, Borderline-SMOTE, VAE, to ten imbalanced datasets. Based on the results, the generation of synthetic samples using VAE yielded better performance. Sun et al. (2020b) reported that VAE performed better than GAN and SMOTE; GAN had better performance compared to SMOTE on some of the datasets. Zhang et al. (2020a) compared random undersampling with Self Organizing Maps (SOM) clustering based undersampling (Vannucci & Colla, 2018). Vannucci & Colla (2018) cluster rare and frequent samples in datasets remove frequent samples to have a more balanced dataset. Zhang et al. (2020a) proposed to use SOM clustering-based undersampling instead of random undersampling. Xu et al. (2019) compared sampling, cost-sensitive, and ensemble learning techniques and found our that cost-sensitive techniques are more effective in improving DL model performance. The experiments conducted by Gong et al. (2019) favored the use of cost-sensitive learning technique over SMOTE for HDP.

## 4.7 DL APPROACHES

As seen in Figure 10, the most frequently used DL algorithm is CNN. The other widely used algorithms are RNN/LSTM/GRU, MLP, and DBN. 13 studies used encode-decoder architecture most of which specifically an Autoencoder.

Like we can observe in some other domains, like malware detection (Catal et al., 2021) and phishing detection (Catal et al., 2022), CNN, RNN/LSTM/GRU, and MLP are the top three most frequently used DL approaches. The overall reason may be that these algorithms performed well in many tasks, and they are well-known among researchers and practitioners; this fact is indeed mentioned in numerous works covered in this study. CNNs, in particular, are reported to work well with high dimensional data and capture local patterns (Li et al. 2017, Pan et al. 2019). In turn, RNN/LSTM/GRU architectures can, for instance, capture long-distance dependencies and semantics (Wang et al. 2021, Liang et al. 2019). DBNs are used with regards to their ability to learn a representation for reconstructing the training data with a high probability (Wang et al. 2016). Finally, autoencoders help learning semantic information and eliminating noise (Zhang et al., 2021).

Less frequently used DL algorithms include Generative Adversarial Networks (GAN), Hybrid DL Model, Hierarchical NN, Siamese NN, Extreme Learning Machines, and Graph NN. Various reasons have been reported for using such approaches, e.g., using GAN for creating synthetic data (Sun et al. 2020), combining supervised and unsupervised techniques in a hybrid architecture (Albahli 2019), using a hierarchical architecture to ensure full extraction of all the semantic features (Wang et al. 2021), using Siamese networks to work with limited data (Zhao et al. 2016), and capturing source code ASTs in graph neural networks (Xu et al. 2021).

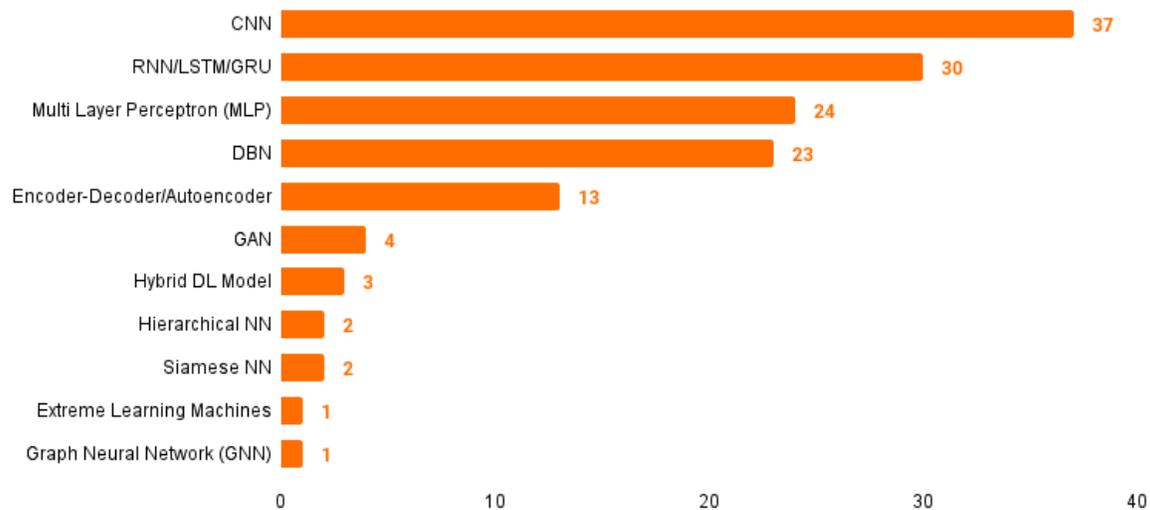

**Figure 10. Distribution of DL approaches**

## 4.8 EVALUATION METRICS AND VALIDATION APPROACHES

Researchers evaluated their defect prediction models using different evaluation metrics and validation approaches. Figure 11 presents the distribution of the top 10 most frequently used evaluation metrics based on our paper pool. 68% of the studies (69) used "F-measure" for evaluation. 48 studies used "Recall", also known as "True Positive Rate" or "Sensitivity". Recall refers to the fraction of the successfully predicted defects. "Area Under the Curve (AUC)" measures a classifier's ability to distinguish between classes. When AUC approaches to one, this means that the prediction model can distinguish positive and negative classes, i.e., buggy and non-buggy in our case. A predictor with an AUC value close to zero tend to classify buggy cases as non-buggy and vice-versa. 40 studies in our paper pool used AUC as an evaluation metric. 40 studies used "Precision", also known as "Positive Predictive Value". Precision refers to the number of correctly predicted defects divided by the number of predictions. The "Accuracy" evaluation metric was used in 28 studies. This metric is easy to understand and easily suits for binary and multi-class classification problems. On the other hand, accuracy metric works well when there is no class imbalance. A defect prediction dataset is qualified as imbalanced when the number of buggy samples much lower than the number of non-buggy ones and vice versa. 20 studies used "Matthews Correlation Coefficient (MCC)" as an effective solution overcoming the class imbalance issue. MCC produces a high score only if the prediction obtained good results in all the four confusion matrix categories (true positives, false negatives, true negatives, and false positives).

Effort-aware defect prediction models consider the differences in the cost of applying quality assurance activities for each piece of code (file, function, etc.) (Mende & Koschke, 2010). Although it may be hard to quantify effort, researchers proposed effort-aware performance measures. 13 studies in our paper pool used an effort-aware performance measure. Six studies, i.e., Albahli (2019), Chen et al. (2019), Qiao & Wang (2019), Qiu et al. (2019c), Wang et al. (2021), and Yedida & Menzies (2021), used "$P_{opt}$" evaluation metric. A larger $P_{opt}$ value refers to a smaller difference between the optimal and predicted models and thus better performing model (Bennin et al., 2016). Four studies, i.e., Sheng et al. (2020), Wang et al. (2020), Zhu et al. (2020), and Wang et al. (2021), used "PofB20" (Jiang et al., 2013) to measure the percentage of defects that a developer can identify by inspecting the top 20% lines of code. Four studies, i.e., Qiao & Wang (2019), Xu et al. (2019), Xu et al. (2021b), and Zhao et al. (2021b), Effort-Aware recall (EARecall), which is defined as the percent of reviewed defective commit instances to the whole defective commit instances. Three studies, i.e., Xu et al. (2019), Xu et al. (2021b), and Zhao et al. (2021b), Effort-Aware F-measure (EAF-measure), which is defined as the weighted harmonic average between EARecall and EAPrecision. Xu et al. (2019) used EAPrecision in addition to EARecall and EAF-measure. Zhao et al. (2019) used "Normalized Expected Cost of Misclassification (NECM)" proposed by Khoshgoftaar & Seliya (2004) to handle the different misclassification costs. To cope with the imbalanced datasets, they used the reciprocal of NECM to strengthen the punishment for the weight of the majority class and reduce the suppression of the weight of the minority class (Zhao et al., 2019). Wang et al. (2021) used IFA, which counts the number of initial false alarms encountered before the first defect is found (Majumder et al., 2022). Since developers will ignore the suggestions if too many false alarms are offered before reporting a defect. Therefore, smaller IFA values are preferred.

Nine studies used "G-measure", which is a trade-off measure that balances "Possibility of detection (PD)" and "Possibility of false alarm (PF)" (Yu et al., 2018). Higher PD and lower PF, thus higher G-measure denote a better prediction model (Yu et al., 2018). Six studies used "ROC curve", also known as "Receiver Operating Characteristics Curve". ROC curve compares the True Positive Rate and the False Positive Rate and helps to determine the trade of between these two characteristics.

Four studies used "Mean Squared Error (MSE)", which measures the average of the squares of the errors. Another set of four studies used "True Negative Rate (TNR)/Specificity", which refers to the ratio of genuinely negative cases that is predicted as negative by model.

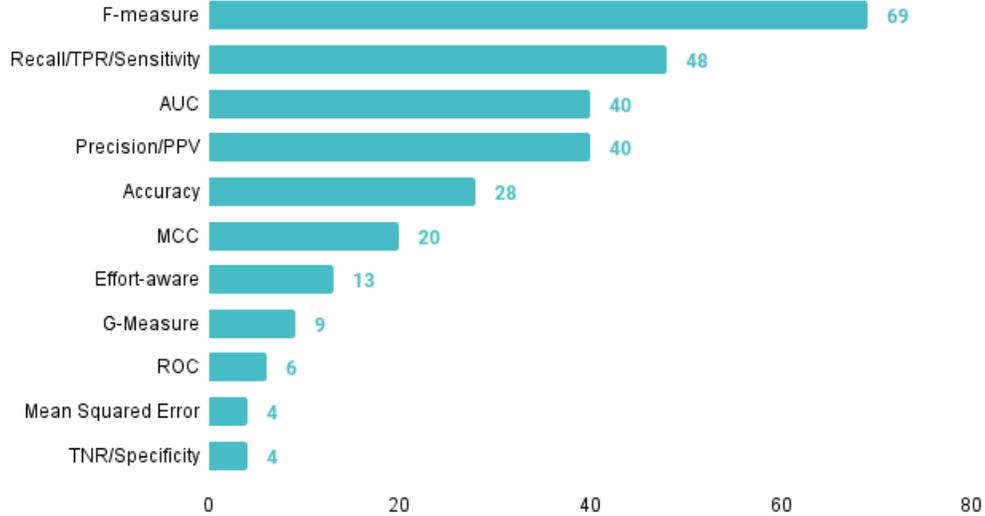

**Figure 11. Distribution of evaluation parameters**

The researchers of 60 studies reported a hold-out approach in which the dataset is split into training and test datasets. Another set of 43 studies preferred cross-validation as a validation approach. Only one study, i.e., Turabieh et al. (2019), reported that both approaches were applied for validation.

## 4.9 Reproducibility Package

The reproducibility of a study is one of the essential characteristics of scientific studies (González-Barahona & Robles, 2012). To be qualified as a reproducible scientific study, the reported experimental results of a study should be obtained by other researchers using authors' artifacts (i.e., source code and datasets) with the same experimental setup (Liu et al., 2021). Some researchers pointed out the reproducibility issues in SE (Lewowski & Madeyski, 2022). Recently Liu et al. (2021) analyzed some studies on the use of DL models in solving a SE problem, like defect prediction or code clone detection. They reported that more than half of the studies do not share high-quality source code or complete data to support the reproducibility of their DL models. Thus, we examined whether the authors of primary studies on SDP using DL publish reproduction packages for their studies. We used the categories used by Lewowski & Madeyski (2022) during data extraction. Figure 12 shows the results on the presence of a reproducibility package in the primary studies in our paper pool. 49 publications (48% of the primary studies) do not mention any sort of package to reproduce the experimental results. Four studies claim to provide a reproducibility package; however, the link in the study is either missing or unresolvable. 36 studies contain only data, and two studies involve only scripts for reproduction. 11 studies (11% of the primary studies) include a reproducibility package including data and scripts.

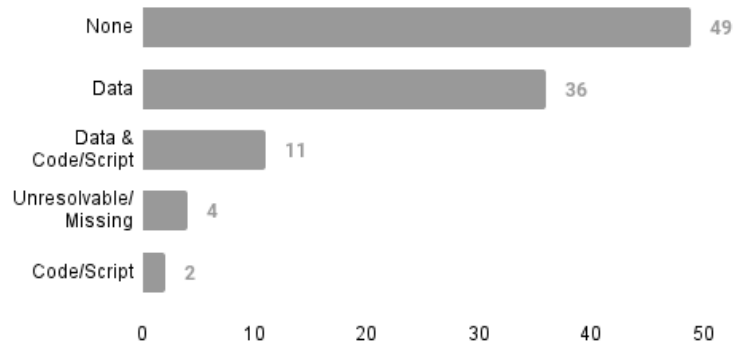

**Figure 12. Presence of reproducibility packages in the primary studies**

In addition, we do not observe an upward trend over the years in sharing a reproducibility package. Out of three and eight studies, only one shared a reproducibility package in 2017 and 2018. In the last three years, out of 33, 26, and 30 studies, only three shared a package for reproduction.

## 4.10 CHALLENGES AND PROPOSED SOLUTIONS

In this section we summarize the challenges and proposed solutions we extracted from 50 primary studies. The rest of the studies do not mention a challenge related to the use of DL for SDP; besides the common challenges we already address explicitly in our research questions (e.g., class imbalance addressed in RQ6). In our qualitative analysis, we mapped the challenges to three categories derived from the ML model life cycle (as introduced in Figure 2):

1. Data Engineering,
2. Model Development, and
3. General (i.e., related to the whole life cycle)

Within each category, we further classified the challenges into subcategories. For each subcategory, we described the challenge, and summarized the solutions offered by the primary studies in the scope of our review. We believe this elaborate analysis, which has not been done with this level of detail in any of the related work (see Section 2.3), can provide researchers with directions to focus their efforts in the coming years.

A short summary of the challenges and their solutions is as follows, while the extensive list can be found in Appendix 8.5.

### 4.10.1 Data Engineering

*Heterogeneous data.* The fact that a large variety of different projects, versions and features is used in SDP leads to highly heterogeneous data, in particular when using different source and target for prediction. Such data degrade the performance of the classifier (Albahli, 2019, Gong et al., 2019; Li et al., 2019a; Qiu et al., 2019a; Qiu et al., 2019b; Sheng et al., 2020; Wang & Lu, 2020; Sun et al., 2020a; Huang et al., 2021; Sun et al., 2021; Wu et al., 2021). Some researchers have tackled this challenge using different DL architectures which take this difference into account, while others have introduced normalization and transformation steps in data preprocessing as well as in feature extraction.

*Insufficient training data.* Having limited training data, either from a quality or quantity point of view, makes it difficult to perform SDP in the first place (Zhao et al., 2018; Zhao et al., 2019; Pandey & Tripathi, 2021). Potential solutions include using DL architectures capable of learning with limited data and adding more labelled data into the training dataset.

*Training data formation.* The training data can contain invalid instances of defects, e.g., incorrectly reported bugs (Li et al., 2019b). A solution to this involves manual validation of the training data, however this has the disadvantage of introducing manual bias in the validation process.

*Incomplete code snippets.* It is a challenge working around incomplete code snippets in change-level SDP, particularly when the approach relies on ASTs of the code (Wang et al., 2020). Heuristic approaches can be used for extracting relevant information from incomplete code.

*AST node granularity and distribution.* Different granularity and distribution for the ASTs used in CPDP might affect SDP negatively (Deng et al., 2020b). A particular multi-kernel transfer CNN, which considers these factors can be used for tackling the challenge.

### 4.10.2 Model Development

*Semantic features:* Traditional features such as source code metrics are not sensitive to the semantics of programs, and can lead to poor prediction performance (Wang et al., 2016; Li et al., 2017; Huo et al., 2018; Cai et al., 2019; Chen et al., 2019; Dam et al., 2019; Fan et al., 2019a; Fan et al., 2019b; Liang et al., 2019; Qiu et al., 2019c; Xu et al., 2019; Wang et al., 2020; Zhang et al., 2020a; Bahaweres et al., 2021, Chatterjee et al., 2021; Xu et al. 2021a; Zhang et al., 2018; Zhang et al., 2021a). A wide range of solutions have been offered to tackle this. These include using richer features ranging from code comments, embeddings, AST, and structural features, to feeding the source code itself to deep learners capable of capturing semantic information and using hybrid and ensemble techniques.

*Feature redundancy:* Highly correlating features may reduce prediction performance (Tran et al., 2019; Turabieh et al., 2019; Wei et al., 2019; Zhao et al., 2021b; Zhu et al., 2021b). Solutions include using deep learning architectures and ensemble techniques which can cope with this and using meta-heuristic approaches for feature selection.

*Manual feature selection:* Chen et al. point out the potential bias due to manual feature selection and rely on an image representation of source code from which image features are automatically selected by a self-attention mechanism (Chen et al., 2020).

*Context-dependence:* Different feature sets may provide the best prediction performance in different contexts, which might be overcome by using a particular LSTM architecture to optimize the combination of input features in each context (Wang et al., 2021).

*Random initialization of parameters:* Random selection of parameters of the learner may lower performance; meta-heuristic approaches can help compute the optimal values (Zhang et al., 2021b).

*Fixed-length feature vectors:* Conventional classification algorithms such as Naïve Bayes, assume the same length for all feature vectors, which can be tackled by using latent features and varying-length features (Wen et al., 2018).

*Sequential networks:* Sequential networks cannot capture the tree syntax and semantics of AST representations and their dependencies (Zhou & Lu, 2020; Yu et al., 2021a). Non-sequential networks such as bi-directional LSTM and HNN can solve the problem.

*Hyperparameter sensitivity:* Hyperparameters are very sensitive for DL models and different settings lead to very different performance results for SDP. Particularly given that not in all cases the hyperparameters are entirely reported, this is a serious challenge in transparency and reproducibility (Pan et al., 2019).

*Model overfitting:* Overfit models over the training data is a potential challenge and can be overcome using suitable techniques such as dropout regularization (Pandey & Tripathi, 2020).

*Performance degradation:* The performance of a prediction model may degrade, which can be avoided by updating the model via user feedback (Albahli, 2019).

### 4.10.3 General

*Early SDP:* Early defect prediction is a challenge; Manjula and Florence suggest combining DNN with generic algorithms for feature optimization (Manjula & Florence, 2019).

*Granularity:* Prediction of too coarse-grained levels of errors burden the developers for bug localization; therefore, keeping SDP at statement-level is a solution (Majd et al., 2020; Munir et al., 2021).

## 5 DISCUSSION

In the following sub-sections, we discuss the results of our study. In Section 5.1, we provide a critical reflection on the results. Section 5.2, we discuss the threats to the validity of the present study and how we addressed them.

### 5.1 GENERAL DISCUSSION

We summarize the key findings to date in terms of current DL approaches and limitations in the SE literature. Additionally, we draw on the findings to provide key recommendations for future research in the application of DL for SDP.

There is a lack of consensus on the evaluation criteria for SDP. Several evaluation criteria have been proposed and each evaluation criteria works very well in specific scenarios. With DL being a subset of ML, it is not unusual that studies evaluate DL models with evaluation indicators used in ML. Popular indicators such as accuracy and precision have been considered as not very feasible in assessing defect prediction models because they are unstable for highly unbalanced datasets (Menzies et al. 2006, Joshi et al. 2001). Menzies et al. (2006) argued that recall and probability of false alarms (*pf*) are good indicators for determining the performance of a SDP model since defect prediction has a challenge, which is highly imbalanced datasets. Additionally, MCC has been recommended to be a better evaluation indicator for highly imbalanced datasets. Our analysis revealed that majority of the studies focused on using F-measure, which computes the harmonic mean between recall and precision and few considered indicators that include *pf*. Recall and AUC were the next most used indicators implying researchers consider the class imbalance issue when assessing the performance of deep learning models. However, additional indicators such as G-measure and MCC are amongst the least used. This could be due in part to the fact that most studies considered less than three evaluation indicators per study and usually used indicators that are easier to compute from a confusion matrix. Although few studies use more than three evaluation indicators in assessing their DL models, the feasibility, understanding and interpretability of these evaluation methods need clearer formalization and empirical studies under different testing scenarios.

We observe that although DL models improved prediction performance, the improvement was not significant when compared to traditional ML models. This may be due in part to the limited amount of data. Our analysis revealed that researchers used the same amount of data they used for conventional ML based defect prediction. Most software projects are not too large in terms of source code files but the data extracted from these projects could be large depending on the metric being used. Metrics computed on function level could generate more data instances compared to metrics computed on file level. Few studies (6%) computed metrics on function or statement level whereas 36% of the studies used file-based metrics. A potential approach to obtaining more metrics, data and improving prediction performance will be to leverage other sources of defect data and not only source code data. Heterogeneous data such as issue tracking documents, bug reports, requirement documents, and test reports among others could be leveraged to produce a massive amount of data for DL models. Devanbu et al. (2020) acknowledge the importance of considering these kind of heterogeneous documents represented in various formats such as source codes, natural language documents, and graphical artifacts as software developers use documents in their daily workflow activities. Documents generated in these formats provide diversity and capture different orthogonal properties and information about the software system, which can inherently aid in improving the performance of deep learning models trained on them. They thus advocate for future practitioners to find several creative methods of combining these information sources for a richer dataset.

WPDP still seems to be the main defect scenario considered by researchers. This is not surprising as software projects are usually developed in versions making it easy to collect larger volumes of data for a single software project. Consequently, the data collected are similar and homogenous, which makes WPDP models perform significantly better than CPDP models. Generally, defect datasets are not that much suitable for training a DL model but rather sufficient only for traditional ML models. CPDP models are trained on different software datasets, which are heterogeneous to the test dataset. However, CPDP is still gaining momentum as our analysis revealed that almost a third of the selected primary studies conducted CPDP. CPDP studies can leverage an advantage of DL, which is that DL models require a lot of data for significantly improved performance. CPDP studies can use several datasets from different software projects for model training, thus making it more appealing than WPDP.

A key advantage of using DL models is the ability to automatically extract features from the data thus eliminating the manual effort of extracting features from data. The review revealed that almost all researchers provide as input either software metrics or convert the source codes in ASTs to the DL models. Two problems with such data inputs are that manually extracted metrics cannot be fully exploited by DL models, such as CNN, which require mostly data in the form of images. Additionally, converting source code into AST implies an additional step of using a tool, which might be somewhat complicated especially for non-Java source code. In addition, the application of DL for SDP is more complex than the traditional ML approaches. Our analysis revealed the complicated techniques and approaches researchers used in extracting features from source codes. For traditional ML approaches, simple tools, which are mostly freely available written in C or Java automatically extract well known metrics from source code. The use of DL for defect prediction requires researchers to develop new tools to convert the source codes to other representations such as AST, control flow and images because proper tools either do not exist or are not widely available. The source code conversion and automatic feature extraction phase remains one of the main challenges stifling the fast progress of the adoption and use of DL for defect prediction. Software data is mostly source code and commit messages, which can be considered as being not very suitable for most DL models. Converting source codes into images looks promising and only one study (Chen et al., 2020) has been able to investigate that possibility. Their study aims to avoid the use of feature extraction tools thus eliminating the use of intermediary representations e.g., ASTs, and instead obtaining code semantic information directly. Thus, they designed a novel, color-based augmentation method to generate 6 color images from each source code file which has been converted into a vector of 8-bit unsigned integers corresponding to the ASCII decimal values of the characters in the source code. Their results were more promising and future research should focus on proposing and designing techniques of converting source code/commit messages into images, which encapsulates the source code information from humans but can be read and processed by DL models. This would also encourage private/commercial organizations to freely provide their source code since researchers would only obtain images without exposing their source code to the public thus ensuring data privacy.

Our analysis revealed that few studies provide replication packages for their experiments. Liu et al. (2021) highlighted the importance of reproducibility and replication of DL studies for SE research. They noted that several studies do not provide their artifacts (source code and datasets) with the same experimental setup, which may be due in part to the complex nature of the experiments, several manual parameters and time-consuming optimization process, which is the opposite of the conventional ML models. Similar to their findings, we observe that very few (11%) provide their source code and dataset to support the reproducibility of their studies. The majority share their dataset, but this is because they all use already existing publicly available datasets. This finding suggests the urgent need to motivate researchers to make their artifacts publicly available. Additionally, researchers who make their complete reproduction packages publicly available should share their packages on well-established research data archives and not their own websites (Lewowski & Madeyski, 2022) since personal websites may not have working links after a period.

There is still little evidence on the practicability and adoption of defect prediction models in industry. Most of the studies used publicly available datasets (Section 4.3) and these models were not evaluated with commercial datasets. This may be due in part to the difficulty of obtaining commercial software data. The difficulty can be linked to data privacy and security, which hinders software quality practitioners in adopting defect prediction models. However, the major challenge to practitioners adopting defect prediction is the complexity of use and incompatibility with their personal or organizational environments. Wan et al (2018) conducted a survey with 395 practitioners and found that as few as 7.8% are willing to use defect prediction tool depicting the low perceived importance of defect prediction by practitioners.

As discussed above (Section 2.1), the process of defect prediction is not very trivial especially regarding data collection and model construction. The use of DL should be targeted to making it easier and more adaptable for practitioners to use. Researchers should consider channeling the immersive power of DL to make defect prediction and detection easier for software quality teams. The popular conventional method of data extraction where source codes are converted to ASTs can be improved by rather converting them into images, which provides several advantages such as encapsulating the data thus contributing to data privacy and the ability to easily provide them to several DL models. Additionally, a systematic DL workflow and taxonomy should be provided to help practitioners comprehend the main difference between defect prediction using DL approaches and defect prediction using traditional ML approaches. The workflow can elaborate on the DL abstractions, process, procedures required for an improved prediction performance.

Few studies considered the application of data resampling approaches. This is a well-known challenge of software defect datasets (Song et al. 2018, Bennin et al, 2019). This is unsurprising as data resampling approaches are usually applied to traditional defect datasets, which are tabular in nature and thus much less complex to pre-process with data resampling. As we observed in Section 4.4, 37% of the studies represented their source codes in an AST format, which may not be feasible for applying data resampling approaches. Nevertheless, the solution is not so far-fetched. Practitioners can leverage already existing data augmentation techniques, which are usually applied to DL datasets to solve class imbalance and aid in data generalization. However, these data augmentation techniques can only be applied to image datasets.

In summary, we provide some key recommendations to address the issues extracted from our findings and discussed above. These recommendations have already been highlighted above and they include:

- Development of new, more comprehensive DL approaches that automatically captures the needed features in sufficient detail and quality from source codes, bug reports and others
- Adoption of data augmentation techniques to tackle the class imbalance issue if the data is efficiently converted into images
- Identification of the key source code defect attributes that need to be captured sufficiently and to support defect prediction
- Publication of replication packages
- Consideration of other sources of data such as requirement documents, test documents, graphical artifacts among others in addition to the source codes.

## 5.2 LIMITATIONS AND POTENTIAL THREATS TO VALIDITY

The scope of this study is limited to the following parameters:

- Date: This study covers primary studies published until the end of 2021, i.e., 31 December 2021.
- Type of Literature: This study comprises studies published in peer-reviewed journals and conference/workshop/symposium proceedings. Grey literature, e.g., papers only published in arxiv.org, blogs, videos, etc., was excluded from the paper pool.
- Perspective: The primary studies were selected using the inclusion criterion of applying at least one DL algorithm to SDP problem and reporting related empirical results. The studies involving only traditional ML algorithms and statistical techniques were excluded.

Some validity considerations are applicable for SLR studies (Petersen et al., 2008; Petersen et al., 2015). The threats to the validity of this study are mainly related to the identification of the candidate pool of papers, primary selection bias, data extraction, and data synthesis.

The selection of search terms and the limitations of search engines can lead to an incomplete set of candidate pool of papers. We carefully selected our search terms by examining related work and queried five widely used online databases used in SLRs on SE topics. We also combined database search with manual inspection of related reviews and forward snowballing using Google Scholar, while there is a risk of missing out some studies due to not performing backward snowballing (Badampudi et al., 2015). Nevertheless, we think that an adequate set of primary studies was collected for this study.

Application of inclusion and exclusion criteria is subject to researchers' bias and hence a potential threat to validity. The authors built a list of inclusion and exclusion criteria and applied a joint voting mechanism to mitigate the risk of ambiguous interpretations. Two authors independently applied inclusion and exclusion criteria to each candidate paper and agreed on 86% of the papers. All the conflicts between two authors' votes were recorded and resolved via the votes of the third and fourth authors.

Another essential aspect that directly affects the results of this study is the validity of the data extraction. The authors started with initial categories that were formed using the existing categories in the literature. In addition, the authors refined the categories iteratively and incrementally. They aimed at decreasing the risk of researcher bias via mapping the relevant data in primary studies to the specified categories. Whenever an author was undecided about the data to be extracted, he recorded that case, and these cases were resolved via discussions among the authors.

We used descriptive statistics to synthesize data for the RQs from one to nine. We think that threats to internal validity are relatively small for the responses to these RQs. We applied the open coding technique iteratively and incrementally to identify the challenges and solutions (RQ10). This coding process potentially entails some researcher bias.

# 6 CONCLUSION AND FUTURE WORK

SDP comprises various techniques for automatically identifying defects and therefore help reduce the effort in fixing them. This is particularly beneficial nowadays given the increasing volume of software and scarce quality assurance resources. SDP using DL has particularly gained traction in the recent years. In this study, we performed a systematic literature review of existing SDP techniques using DL to paint a picture on the state-of-the-art. We applied a rigorous process to search for articles in several scientific databases, supported with snowballing. As a result of a multiple-assessor quality assessment step with well-defined criteria, we chose the articles to be considered for analysis. Our survey eventually included a total of 102 high-quality primary studies. Based on those we conducted quantitative and qualitative analysis on the pool of studies with respect to various aspects of SDP: SDP scenarios, ML categories, datasets, representation of source code, granularity level of prediction, dealing with the class imbalance problem, DL approaches, evaluation metrics and validation approach, reproducibility, and finally challenges along with several proposed solutions.

The results indicate an increasing trend of SDP research over the recent years, with a big variety of fundamental techniques, datasets and validation approaches being employed. An important observation is the lack of reproducibility packages for most of the surveyed articles, which can be problematic for transparency and further advancement of the field. We have also collected the reported challenges around the data engineering, model development aspects and SDP in general, along with several solutions proposed by researchers. Together with our critical discussion, we propose the following directions to pave the way for further research:

- Development of new, more comprehensive DL approaches automatically capturing richer representations and features from heterogeneous sources (source code, bug reports and others),
- Development of data augmentation techniques for tackling limited dataset sizes and class imbalance,
- Identification of key source code defect attributes for defect prediction as well as exploitation of automatic feature extraction of DL approaches,
- Establishing common criteria for evaluating the performance of DL-based SDP,
- More focus on CPDP scenarios next to WPDP,
- Better usability of SDP tools and integration into the daily practice of users,
- Reproducibility and open science.

Our results can be beneficial for both newcomers to SDP research to see the landscape of different approaches, and established researchers to focus their efforts in the coming years. As future work we aim to perform a more in-depth investigation into the state-of-the-art in SDP using DL, particularly doing an extensive meta-analysis on factors influencing the performance of SDP as reported in the surveyed articles in this study.

# 8 APPENDICES

## 8.1 SEARCH STRINGS FOR ONLINE DATABASES

| ACM | The below query was executed using "edit query" feature on advanced search interface. |
|---|---|
| | *((Title: software) AND ((Title: fault) OR (Title: defect) OR (Title: quality) OR (Title: bug)) AND ((Title: predict\*) OR (Title: estimat\*)) AND (Title: "deep learning")) OR ((Abstract: software) AND ((Abstract: fault) OR (Abstract: defect) OR (Abstract: quality) OR (Abstract: bug)) AND ((Abstract: predict\*) OR (Abstract: estimat\*)) AND (Abstract: "deep learning")) OR ((Keyword: software) AND ((Keyword: fault) OR (Keyword: defect) OR (Keyword: quality) OR (Keyword: bug)) AND ((Keyword: predict\*) OR (Keyword: estimat\*)) AND (Keyword: "deep learning"))* |
| IEEE Xplore | The below query was executed using "command search" feature on advanced search interface. |
| | *((("Document Title":software) AND ("Document Title":fault OR "Document Title":defect OR "Document Title":bug OR "Document Title":quality) AND ("Document Title":predict\* OR "Document Title":estimat\*) AND ("Document Title":"deep learning")) OR (("Abstract":software) AND ("Abstract":fault OR "Abstract":defect OR "Abstract":bug OR "Abstract":quality) AND ("Abstract":predict\* OR "Abstract":estimat\*) AND ("Abstract":"deep learning")) OR (("Author Keywords":software) AND ("Author Keywords":fault OR "Author Keywords":defect OR "Author Keywords":bug OR "Author Keywords":quality) AND ("Author Keywords":predict\* OR "Author Keywords":estimat\*) AND ("Author Keywords":"deep learning")))* |
| ScienceDirect | The two queries below were executed using advanced search interface. The reason of using two separate queries was that the search feature did not allow the use of wildcard (\*). The results were combined, and the duplicates were removed. |
| | Query 1: *(software) AND (fault OR defect OR bug OR quality) AND (predict OR prediction) AND ("deep learning")* |
| | Query 2: *(software) AND (fault OR defect OR bug OR quality) AND (estimate OR estimation) AND ("deep learning")* |
| Springer | The eight queries below were executed using search interface. The content types "book", "protocol", "reference work" were removed from the results. The results were combined, and the duplicates were removed. |
| | Query 1: *"software fault" estimate\* "deep learning"* |
| | Query 2: *"software fault" predict\* "deep learning"* |
| | Query 3: *"software defect" estimate\* "deep learning"* |
| | Query 4: *"software defect" predict\* "deep learning"* |
| | Query 5: *"software bug" estimate\* "deep learning"* |
| | Query 6: *"software bug" predict\* "deep learning"* |
| | Query 7: *"software quality" estimate\* "deep learning"* |
| | Query 8: *"software quality" predict\* "deep learning"* |
| Wiley | The below query was executed three times by selecting "Title", "Abstract", and "Keywords" as the search field on advanced search interface. The results were combined, and the duplicates were removed. |
| | *(software) AND (fault OR defect OR bug OR quality) AND (estimat\* OR predict\*) AND ("deep learning")* |

## 8.2 LIST OF EXCLUDED STUDIES

We excluded the below listed 14 studies since they did not meet our quality assessment criteria.

## 8.3 DISTRIBUTION OF THE PRIMARY STUDIES PER VENUE

| Venue | Number of Primary Studies | Reference(s) |
|---|---|---|
| IEEE Access | 10 | Al Qasem et al. (2020), Cai et al. (2019), Chen et al. (2019), Deng et al. (2020b), Liang et al. (2019), Lin & Lu (2021), Qiu et al. (2019c), Sheng et al. (2020), Sun et al. (2020a), Zhao et al. (2018) |
| IET Software | 5 | Deng et al. (2020a), Huang et al. (2021), Zhang et al. (2021b), Zhao et al. (2021b), Zhu et al. (2020) |
| International Conference on Software Quality, Reliability and Security (QRS) | 4 | Li et al. (2017), Yang et al. (2015), Zhang et al. (2018), Zhou & Lu (2020) |
| Expert Systems with Applications | 3 | Majd et al. (2020), Pandey et al. (2020), Turabieh et al. (2019) |
| IEEE Transactions on Reliability | 3 | Wang et al. (2021), Xu et al. (2021a), Xu et al. (2021b) |
| IEEE Transactions on Software Engineering | 3 | Yedida & Menzies (2021), Wang et al. (2018), Wen et al. (2020) |
| Applied Sciences | 2 | Pan et al. (2019), Qiu et al. (2019b) |
| Asia-Pacific Software Engineering Conference (APSEC) | 2 | Fan et al. (2019b), Zhang et al. (2020a) |
| Information and Software Technology | 2 | Tong et al. (2018), Zhou et al. (2019) |
| Information Sciences | 2 | Zhang et al. (2021a), Zhu et al. (2021a) |
| International Conference on Dependable Systems and Their Applications (DSA) | 2 | Liu et al. (2020), Yu et al. (2021b) |
| International Conference on Mining Software Repositories (MSR) | 2 | Dam et al. (2019), Hoang et al. (2019) |
| International Conference on Software Engineering (ICSE) | 2 | Chen et al. (2020), Wang et al. (2016) |
| International Conference on Software Engineering and Knowledge Engineering | 2 | Qiu et al. (2019a), Wang & Lu (2020) |
| International Workshop on Realizing Artificial Intelligence Synergies in Software Engineering (RAISE) | 2 | Humphreys & Dam (2019), Young et al. (2018) |
| Journal of Systems and Software | 2 | Xu et al. (2019), Zhu et al. (2021b) |
| Knowledge-Based Systems | 2 | Pandey & Tripathi (2020), Pandey & Tripathi (2021) |
| Neural Computing and Applications | 2 | Ardimento et al. (2021), Nevendra & Singh (2021) |
| Neurocomputing | 2 | Qiao et al. (2020), Zhao et al. (2019) |
| PLOS ONE | 2 | Munir et al. (2021), Qiao & Wang (2019) |
| ACM on Programming Languages (OOPSLA) | 1 | Li et al. (2019c) |
| ACM SIGSOFT International Symposium on Software Testing and Analysis | 1 | Zeng et al. (2021) |
| Algorithms and Architectures for Parallel Processing | 1 | Sun et al. (2020b) |
| Annual ACM Symposium on Applied Computing | 1 | Zhao et al. (2021a) |
| Array | 1 | Ferenc et al. (2020) |
| Asia Pacific Symposium on Intelligent and Evolutionary Systems (IES) | 1 | Phan & Le Nguyen (2017) |
| Chinese Conference on Pattern Recognition and Computer Vision (PRCV) | 1 | Li et al. (2019b) |
| Chinese Journal Electronics | 1 | Wei et al. (2019) |
| Cluster Computing | 1 | Manjula & Florence (2019) |
| Cognitive Systems Research | 1 | Geng (2018) |
| Conference on Knowledge Based Engineering and Innovation (KBEI) | 1 | Eivazpour & Keyvanpour (2019) |
| Congress on Intelligent Systems | 1 | Thaher & Khamayseh (2021) |
| Euromicro Conference on Software Engineering and Advanced Applications (SEAA) | 1 | Fiore et al. (2021) |
| Future Internet | 1 | Albahli (2019) |
| IEEE Annual Computers, Software, and Applications Conference (COMPSAC) | 1 | Yu et al. (2021a) |
| IEEE International Conference for Innovation in Technology (INOCON) | 1 | Yadav (2020) |
| IEEE International Conference on Big Data and Cloud Computing (BdCloud) | 1 | Sun et al. (2018) |
| IEEE International Conference on Data Mining (ICDM) | 1 | Huo et al. (2018) |
| IEEE International Conference on Industrial Engineering and Engineering Management (IEEM) | 1 | Huang et al. (2019) |
| IEEE Uttar Pradesh Section International Conference on Electrical, Electronics and Computer Engineering (UPCON) | 1 | Bhandari & Gupta (2018) |
| IEICE TRANSACTIONS on Information and Systems | 1 | Gong et al. (2019) |
| Intelligent Data Analysis | 1 | Saifan & Al Smadi (2019) |
| International Conference on Advances in Science, Engineering and Robotics Technology (ICASERT) | 1 | Ayon (2019) |
| International Conference on Advances in the Emerging Computing Technologies (AECT) | 1 | Abozeed et al. (2020) |
| International Conference on Artificial Intelligence and Security (ICAIS) | 1 | Sun et al. (2021) |
| International Conference on Artificial Intelligence for Communications and Networks (AICON) | 1 | Zhu et al. (2019) |
| International Conference on Computational Performance Evaluation (ComPE) | 1 | Singh et al. (2020) |
| International Conference on Computer Communications and Networks (ICCCN) | 1 | Tian & Tian (2020) |
| International Conference on Computer Systems and Applications (AICCSA) | 1 | Samir et al. (2019) |
| International Conference on Electrical Engineering, Computer Sciences and Informatics (EECSI) | 1 | Bahaweres et al. (2020) |
| International Conference on Internet of Things and Connected Technologies (ICIoTCT) | 1 | Chatterjee et al. (2021) |
| International Conference on Knowledge and Systems Engineering (KSE) | 1 | Tran et al. (2019) |
| International Conference On Smart Cities, Automation & Intelligent Computing Systems (ICON-SONICS) | 1 | Bahaweres et al. (2021) |
| International Conference on Soft Computing and Signal Processing (ICSCSP) | 1 | Malohtra & Yadav (2021) |
| International Conference on Tools with Artificial Intelligence (ICTAI) | 1 | Phan et al. (2017) |
| International Cyberspace Congress, CyberDI and CyberLife | 1 | Yang et al. (2019) |
| International Joint Conference on Neural Networks (IJCNN) | 1 | Li et al. (2019a) |
| International Journal of Web Services Research (IJWSR) | 1 | Bhandari & Gupta (2020) |
| Journal of Computer Languages | 1 | Shi, et al. (2020) |
| Journal of Software: Evolution and Process | 1 | Shi, et al. (2021) |
| Mathematical Problems in Engineering | 1 | Song et al. (2021) |
| PeerJ Computer Science | 1 | Farid et al. (2021) |
| Progress in Advanced Computing and Intelligent Engineering (ICACIE) | 1 | Tameswar et al. (2021) |
| Scientific Programming | 1 | Fan et al. (2019a) |
| Software Quality Journal | 1 | Wu et al. (2021) |
| Wireless Personal Communications | 1 | Dong et al. (2018) |

## 8.4 PRIMARY STUDIES (SOURCES REVIEWED IN THE SLR)

## 8.5 THE CHALLENGES AND PROPOSED SOLUTIONS

| Category | Challenge | Proposed Solution |
|---|---|---|
| Data Engineering | Source and target dataset's different features and data distributions degrade the performance of classifier (Albahli, 2019, Gong et al., 2019; Li et al., 2019a; Qiu et al., 2019a; Qiu et al., 2019b; Sheng et al., 2020; Wang & Lu, 2020; Sun et al., 2020a; Huang et al., 2021; Sun et al., 2021; Wu et al., 2021) | Li et al. (2019a) used a cost-sensitive shared hidden layer autoencoder with shared parameter mechanism to make the distribution of source and target datasets more similar by minimizing reconstruction error loss. |
| | | Qiu et al. (2019a) proposed a Transferable Hybrid Features Learning with CNN. |
| | | Qiu et al. (2019) employed a matching layer to bridge the source and target datasets to mine the transferable semantic-based features by simultaneously minimizing classification error and distribution divergence between projects. |
| | | Albahli (2019) checked training data against outliers and processed these outliers accordingly to obtain a better model. |
| | | Wang & Lu (2020) introduced a domain confusion loss based maximum mean discrepancy (MMD) in feature extraction to bridge the substantial distributional discrepancy between different projects. |
| | | Wu et al. (2021) adopted a modified autoencoder algorithm for instance selection. |
| | | Sheng et al. (2020) proposed Adversarial Discriminative Convolutional Neural Network (ADCNN) to extract transferrable semantic features from source code for CPDP tasks. |
| | | Sun et al. (2021) proposed a deep adversarial learning based HDP approach and leveraged DNN to learn nonlinear transformation for each project to obtain common feature representation, which the heterogeneous data from different projects can be compared directly. |
| | | Huang et al. (2021) proposed a model based on multi-adaptation and nuclear form to deal with different samples. |

| | | Gong et al. (2019) utilized the Maximum Mean Discrepancy (MMD) to calculate the distance between the source and target data. |
|---|---|---|
| Data Engineering | Developing a successful SDP model is a challenge when training data with sufficient amount and quality are not present (Zhao et al., 2018; Zhao et al., 2019; Pandey & Tripathi, 2021). | Zhao et al. (2018) and Zhao et al. (2019) proposed using Siamese networks, which are capable of learning with a few samples.<br><br>Pandey & Tripathi (2021) added more labelled data to their training set to improve the performance of DNN-based model. |
| Data Engineering | Training data formation (Li et al., 2019b) | Li et al. (2019b) manually checked bug reports to validate whether reported bugs were true bugs. On the other hand, this introduces a bias since researchers are not the actual developers of the projects and hence may misunderstand the code and bug. |
| Data Engineering | For change-level defect prediction, code snippets are used as training data and building AST for an incomplete code snippet is challenging (Wang et al., 2020). | Wang et al. (2020) proposed a heuristic approach to extracting important structural and context information from incomplete code snippets. |
| Data Engineering | In a CPDP scenario, the granularity of the AST nodes and the data distribution difference among datasets may have negative impacts on the prediction performance (Deng et al., 2020b). | Deng et al. (2020b) proposed a CPDP framework based on multi-kernel transfer CNNs by considering AST node granularity. |
| Model Development | Traditional features (such as lines of code, operand and operator counts, number of methods in a class, the position of a class in inheritance tree, and McCabe complexity) are not sensitive to programs' semantic information and hence harm defect prediction performance (Wang et al., 2016; Li et al., 2017; Huo et al., 2018; Cai et al., 2019; Chen et al., 2019; Dam et al., 2019; Fan et al., 2019a; Fan et al., 2019b; Liang et al., 2019; Qiu et al., 2019c; Xu et al., 2019; Wang et al., 2020; Zhang et al., 2020a; Bahaweres et al., 2021; Chatterjee et al., 2021; Xu et al. 2021a; Zhang et al., 2018; Zhang et al., 2021a). | Wang et al., 2016 and Wang et al., 2020 leveraged the semantic features learned by DBN to improve WPDP and CPDP performance.<br><br>Chen et al. (2019) a simplified AST for representation to capture semantic information of source code. They simplified AST by including project-independent nodes and ignoring project specific nodes (such as method and variable names).<br><br>Xu et al. (2021a) used graph neural networks to capture semantic and context information using ASTs and learn latent defect information of defective subtrees.<br><br>Chatterjee et al. (2021) utilized DNN to learn features automatically instead of designing handcrafted features.<br><br>Huo et al. (2018) used code comments to generate semantic features besides other features to train a CNN.<br><br>Dam et al. (2019) used a tree LSTM network that matches with AST representation to represent syntax and semantics of source code better.<br><br>Liang et al. (2019) proposed a Semantic LSTM network to capture semantic information of source code.<br><br>Zhang et al. (2020a) proposed a model based on ensemble learning techniques and attention mechanisms for better source code representation. |

|  |  | Li et al. (2017) used CNN to learn semantic and structural features of programs. |
|  |  | Zhang et al. (2021a) used a hybrid model based on WGAN-GP (Wasserstein GAN with Gradient Penalty), multi-objective NSGA-III (Non-dominated Sorting Genetic Algorithm - III) algorithm and hybrid CNNSVM (Convolutional Neural Network – Support Vector Machine) to represent complex structure of programs. |
|  |  | Xu et al. (2019) use a DNN with a new hybrid loss function that consists of a triplet loss to learn a more discriminative feature representation of defect data. |
|  |  | Qiu et al. (2019c) proposed a new model, named neural forest (NF), which uses the DNN and decision forest to build a holistic system for the automatic exploration of powerful feature representations. |
|  |  | Bahaweres et al. (2021) used AST nodes and word embeddings to build an LSTM network. |
|  |  | Fan et al. (2019a) and Fan et al. (2019b) used word embeddings obtained via ASTs to form numeric vectors. |
|  |  | Zhang et al. (2018) used cross-entropy, a common measure for natural language, as a new code metric and combined it with traditional metrics. |
|  |  | Cai et al. (2019) used an AST-based representation along with Euclidean distance to represent semantic distance between nodes. |
| Model Development | Feature redundancy, i.e., highly correlated features, may harm prediction performance (Tran et al., 2019; Turabieh et al., 2019; Wei et al., 2019; Zhao et al., 2021b; Zhu et al., 2021b) | Tran et al. (2019) leveraged DL and ensemble learning to learn effective representations of metrics. |
|  |  | Wei et al. (2019) used DBN to learn features. |
|  |  | Zhao et al. (2021b) used Principal Component Analysis (PCA) for feature representation learning. |
|  |  | Turabieh et al. (2019) used a pool of meta-heuristic-based feature selection methods (i.e., Genetic Algorithm, Particle Swarm Optimization, and Ant Colony Optimization) to select features. |
|  |  | Zhu et al. (2021b) leveraged Whale Optimization Algorithm (WOA) and another complementary Simulated Annealing (SA) to construct an enhanced metaheuristic search-based feature selection algorithm. |
| Model Development | Manual feature selection may harm prediction performance (Chen et al., 2020) | Chen et al. (2020) represented source code as images, applied self-attention mechanism to extract image features, and fed images to a pre-trained DL model for SDP. |
| Model Development | Different feature sets may provide the best prediction performance in different contexts (Wang et al., 2021). | Wang et al. (2021) used a gated merge layer in their LSTM network to obtain an optimum combination of the input features. |
| Model Development | Random selection of initial input weights and hidden layer biases of Extreme Learning Machine (ELM) may lead to lower model performance (Zhang et al., 2021b). | Zhang et al. (2021b) utilized metaheuristic intelligence optimization algorithms to determine optimal input weights and hidden layer biases of ELM, including Gravitational Search Algorithm (GSA) and Particle Swarm Optimization (PSO). |
| Model Development | Conventional classification algorithms, e.g., Naïve Bayes, Decision Tree, Logistic Regression, assume features represented by vectors of the same length. This scheme is not appropriate for the representation of change sequences with varying lengths (Wen et al., 2018). | Wen et al. (2018) used RNN, which can automatically derive latent features from sequence data. |
| Model Development | Sequential networks do not represent syntax and semantics of AST and fail to capture dependencies in source code (Zhou & Lu, 2020; Yu et al., 2021a). | Zhou & Lu (2020) used a bidirectional LSTM to represent dependencies in source code and tree LSTM network to capture syntactic information from AST.<br><br>Yu et al. (2021a) used a hierarchical neural network. |
| Model Development | DL models are sensitive to hyperparameters leading to very different performance results. This is a serious challenge in reproducing previous experiments for which all hyperparameters were not reported (Pan et al., 2019). | No solution proposed |
| Model Development | Model overfitting (Pandey & Tripathi, 2020) | Pandey & Tripathi (2020) used dropout regularization to avoid overfitting. |
| Model Development | The performance of a prediction model may degrade (Albahli, 2019). | Albahli (2019) adjusted their prediction model according to the feedback (input from users on whether a prediction is correct). |
| General | Early defect prediction is a challenging task (Manjula & Florence, 2019). | Manjula & Florence (2019) combined genetic algorithm for feature optimization with DNN for classification and observed a performance improvement due to the application of optimization technique. |
| General | SDP at course-grained levels puts burden on developers for bug localization (Majd et al., 2020; Munir et al., 2021). | Majd et al. (2020) and Munir et al. (2021) proposed a DL-based approach for statement-level SDP. |